\documentclass[twocolumn]{aastex62}

\pdfoutput=1

\usepackage{amssymb}
\usepackage{amsmath}
\usepackage{latexsym}
\usepackage{natbib}
\usepackage{wasysym}
\usepackage{mathtools}
\usepackage[hang,flushmargin]{footmisc} 
\newcommand{\kep}{\emph{Kepler}\xspace}
\newcommand{\sigz}{\Sigma_{z,1}}
\newcommand{\lesseroneslope}{4\xspace}
\newcommand{\lesserone}{1\xspace}
\newcommand{\ngdtwofive}{5\xspace}
\newcommand{\lessmany}{2\xspace}
\newcommand{\lessertwo}{3\xspace}
\usepackage{xspace}
\usepackage{graphicx}
\usepackage{epstopdf}

\DeclareMathAlphabet{\mathpzc}{OT1}{pzc}{m}{it}

\begin{document}

\pagenumbering{arabic}

\shorttitle{Forming Diverse Super-Earths}
\shortauthors{MacDonald et al.}

\correspondingauthor{Mariah G. MacDonald}
\email{mmacdonald@psu.edu}

\title{Forming Diverse Super-Earth Systems in Situ}
\author[0000-0003-2372-1364]{Mariah G. MacDonald}
\affiliation{Department of Astronomy \& Astrophysics, Center for Exoplanets and Habitable Worlds, The Pennsylvania State University, University Park, PA 16802, USA}
\author[0000-0001-9677-1296]{Rebekah I. Dawson}
\affiliation{Department of Astronomy \& Astrophysics, Center for Exoplanets and Habitable Worlds, The Pennsylvania State University, University Park, PA 16802, USA}
\author[0000-0002-2432-833X]{Sarah J. Morrison}
\affiliation{Department of Physics, Astronomy, \& Materials Science, Missouri State University, Springfield, MO 65897, USA}
\affiliation{Department of Astronomy \& Astrophysics, Center for Exoplanets and Habitable Worlds, The Pennsylvania State University, University Park, PA 16802, USA}
\author[0000-0002-1228-9820]{Eve J. Lee}
\affiliation{Department of Physics and McGill Space Institute, McGill University, 3550 rue University, Montreal, QC, H3A 2T8, Canada}
\author[0000-0002-3791-3650]{Arjun Khandelwal}
\affiliation{Haverford College, 370 Lancaster Avenue, Haverford, PA 19041, USA}

\setcounter{footnote}{0}

\begin{abstract} 
Super-Earths and mini-Neptunes exhibit great diversity in their compositional and orbital properties. Their bulk densities span a large range, from those dense enough to be purely rocky to those needing a substantial contribution from volatiles to their volumes. Their orbital configurations range from compact, circular multi-transiting systems like Kepler-11 to systems like our Solar System's terrestrial planets with wider spacings and  modest but significant eccentricities and mutual inclinations. Here we investigate whether a continuum of formation conditions resulting from variation in the amount of solids available in the inner disk can account for the diversity of orbital and compositional properties observed for super Earths, including the apparent dichotomy between single transiting and multiple transiting system. We simulate in situ formation of super-Earths via giant impacts and compare to the observed Kepler sample. We find that intrinsic variations among disks in the amount of solids available for in situ formation can account for the orbital and compositional diversity observed among Kepler's transiting planets. Our simulations can account for the planets' distributions of orbital period ratios, transit duration ratios, and transit multiplicity; higher eccentricities for single than multi transiting planets; smaller eccentricities for larger planets; scatter in the mass-radius relation, including lower densities for planets with masses measured with TTVs than RVs; and similarity in planets' sizes and spacings within each system. Our findings support the theory that variation among super-Earth and mini-Neptune properties is primarily locked in by different in situ formation conditions, rather than arising stochastically through subsequent evolution.
\end{abstract}


\section{Introduction}
\label{sec:intro}
The astronomical community was surprised to find thousands of planets and planetary candidates between the sizes of Earth and Neptune, now called ``super-Earths,\footnote{Low-density super-Earths are sometimes termed ``mini-Neptunes;" here we use the term super-Earth for all planets in this size range, regardless of composition.}"  which are unlike any planet in our Solar System in mass and size \citep[e.g.,][]{Howard2010,Borucki2011,Batalha2013,Burke2014,Mullally2015}. Super-Earths span a large range of bulk densities, from those dense enough to be purely rocky to those needing a substantial contribution from volatiles to their volumes \citep[e.g.,][]{Weiss2014,Welsh2015,Dressing2015,Wolfgang2016}. 
Super-Earths' planetary systems exhibit a large range of orbital properties, from compact, circular multi-transiting systems like Kepler-11 \citep{Lissauer2011} to systems like our Solar System's terrestrial planets with wider spacings and modest but significant eccentricities and mutual inclinations. In analogy to the classical Kuiper belt, we refer to these as ``dynamically cold'' and ``dynamically hot'' respectively. The so-called ``\kep dichotomy'' between systems with one vs. multiple transiting planets (e.g., \citealt{Lissauer2011}) is one manifestation of this orbital diversity. It is debated which mechanism(s) are predominately responsible for the diversity of orbits and compositions. Possible contributors (which can be complementary rather than being mutually exclusive) include dynamical evolution over time (e.g., \citealt{Pu2015,Volk2015}), presence or absence of a giant planet (e.g., \citealt{Huang2017}), host star obliquity \citep{Spalding2016}, resonant chains that remain stable or disrupt post migration \citep{izid17}, multiple formation channels (e.g.,\citealt{Lee2016}), atmospheric loss (e.g., \citealt{inam15}), or variation in formation conditions (e.g., \citealt{Dawson2015}; \citealt{Dawson2016}, DLC16 hereafter; \citealt{Moriarty2016}, MB16 hereafter).

Here we explore whether an intrinsic variety of disk conditions for in situ formation (e.g., \citealt{Chiang2013,Hansen2013}) can explain the observed orbital and compositional diversity of \kep super-Earths. We previously showed that the amount of solids present in the formation region of the proto-planetary disks strongly influences whether a super-Earth of a given mass forms rocky or accretes a low mass gaseous atmosphere \citep{Dawson2015}. We also found that the amount of gas present at the late stages of formation affects the final orbital properties and links them to compositional properties (DLC16). MB16 demonstrated that both the amount of solids present and their radial distribution affect the final orbital properties of super-Earth systems formed in situ. However, these earlier studies had a number of limitations and ambiguities -- which we will address here -- in exploring whether formation conditions can establish the variety of planets we observe.

First, DLC16 and MB16 each mixed two types of formation conditions to match super-Earths' observed orbital properties. DLC16 mixed planets formed in disks with two different gas conditions and MB16 mixed planets formed in disk with two different radial solid profiles. Here we choose to instead focus on a continuum of the amount of solids present in the formation region because we have reason to suspect variation in this property. In situ formation of close-in super-Earths requires a larger reservoir of solids than building our Solar System's terrestrial planets 
\citep[e.g.,][]{Chiang2013,Schlichting14}
so intrinsic variation in the amount of solids explains why some proto-planetary disks form meager Solar Systems and others form hefty Kepler-11s. The variation in the amount of solids could reflect both the overall disk metallicity and the efficiency of radially transporting and concentrating solids in the inner disk.

Second, DLC16's results were ambiguous about whether differences in gas disk dispersal are necessary to explain the diversity of orbital properties. The two types of formation conditions mixed by DLC16 were two levels of gas in the proto-planetary disk before its dispersal. These two levels corresponded to two qualitatively different super-Earth formation environments: a gas-rich mode in which super-Earths reach their final masses and orbits in the presence of disk gas, and a gas-poor mode in which super-Earths finish assembling after the gas disk dissipates. However, DLC16 pointed out that a single level of disk gas with different amounts of solids could also lead to this diversity of formation environments. There could be intrinsic diversity among disks in both these parameters. Disks \emph{might} vary in their gas levels prior to dispersal in the photo-evaporative switch model of gas disk dispersal (e.g., \citealt{owen11,owen12}) due to different X-ray/FUV stellar luminosity. But -- under the hypothesis of in situ formation -- disks \emph{must} vary in their solid reservoirs close to the star to explain the wide range of observed masses (e.g., \citealt{Chiang2013}).   Models of gas accretion also report the final envelope mass fraction---which is the most dominant proxy of the radii \citep{Lopez14}---of planets to be remarkably insensitive to the nebular gas density \citep[e.g.,][]{Lee2016}. Given the more compelling need, we choose to explore how much of the observed variation in super-Earth orbital and compositional properties we can explain with a continuum in the amount of available solids.

We will also address a few other limitations of earlier studies. DLC16 applied a simple mass and period cut for detection efficiency when comparing simulated planets to observed planets, but assumptions about detection efficiency can affect how many single vs. multi transiting systems our model produces. Here we will forward model detection efficiency based on studies of the \kep pipeline (e.g., \citealt{Christiansen2015,chri16}).  DLC16 failed to reproduce the most compact pairs (i.e. the smallest period ratios for adjacent planets), and we will explore whether this failure is a fundamental limitation of in situ formation or can be addressed with a more flexible range of initial conditions. Finally, we will compare our simulations to more recent observed trends of super-Earth diversity: intrinsic scatter in super-Earths' mass-radius relation \citep{Wolfgang2016} and similarity in size among planets in the same system \citep{mill17,weis18}.

Here we focus on in situ formation -- without migration -- to explore whether the continuum in the amount of available solids can fully account for the diversity of observed properties, without the need to invoke multiple origins channels. In Section~\ref{sec:simulations} we explain our parameterization of disk properties, describe our planet formation simulations, and detail how we compare our simulations to observed \kep planet candidates. We explore how a continuum of in situ solid reservoirs can lead to diversity in their orbits (Section~\ref{sec:orbits}) and compositions (Section~\ref{sec:compositions}). We summarize our results in Section~\ref{sec:conclusion}.

\section{Simulations and Observables}\label{sec:simulations}

Here we give an overview of the planet formation scenario we simulated and how we compare the simulated planets to observed \kep planet candidates. We model in situ formation of super-Earths in a proto-planetary disk. While the gas disk is fully present, planetary embryos grow from a reservoir of solid material that we assume was transported from the outer disk as dust, planetesimals, and/or smaller embryos. As the gas disk dissipates, the embryos become less gravitationally cushioned from each other and start to merge and grow via giant impacts. The planets accrete gas, and the gas continues to affect their orbits. The disk gas surface density gradually declines until it reaches a critical value and then quickly vanishes (e.g., \citealt{owen11,owen12}). After the gas disk stage, the planets continue to interact gravitationally -- sometimes colliding -- over billions of years. Depending on how each planetary system is orientated on the sky and the detectability of its planets, the \kep spacecraft might observe transits of one or more planet.

We employ a number of tools and assumptions to approximate this scenario. In Section \ref{subsec:disk}, we describe the parameterization of disk properties. In Section \ref{subsec:sims}, we detail our simulations. We explain how we compare to \kep planet candidates in Section \ref{subsec:compare}.

\subsection{Disk properties}
\label{subsec:disk}

Here we describe our parameterization of the local disk properties during planet formation. We assume the reservoir of solids is radially distributed according to the surface density
\begin{equation}\label{eq:alpha}
    \Sigma_z = \sigz\bigg(\frac{a}{\textrm{au}}\bigg)^{\alpha}
\end{equation}
\noindent where $a$ is the semi-major axis and $\sigz$ is the solid surface density at 1 AU. Often, $\alpha$ is taken to be -1.5 (e.g., \citealt{cham01}), as in the minimum mass solar nebula. Because of the flexibility in choosing $\sigz$ and $\alpha$, we can remain somewhat agnostic about which physical processes delivered and distributed the reservoir of solids. The biggest assumption is that the distribution of solids is smooth. In future work, it would be interesting to explore a disk with pile-ups and gaps in the distribution of solids.

We assume that the gas surface density follows
\begin{equation}\label{eq:gas}
    \Sigma_g = 1700{~\rm g cm}^{-2} d^{-1}  \bigg(\frac{a}{\textrm{au}}\bigg)^{-1.5}
\end{equation}
\noindent where the depletion factor $d=1$ corresponds to the minimum-mass solar nebula and $d>1$ corresponds to a more depleted nebula. The scenario we envision is that $1/d$ declines gradually over time until it reaches some threshold value and then quickly declines to $1/d=0$ (the photoevaporative switch model).  However, since our simulations begin at the embryo stage, here we approximate the dissipation process as a step function: we begin with $d$ at its threshold value for a 1 Myr, and subsequently $d=0$. This timescale should be interpreted as the dissipation timescale at the end of the disk lifetime rather than the full disk lifetime. Because we simulate the end of the disk lifetime, we assume that the solid material is largely in place at the beginning of the simulation, rather than being delivered throughout -- or partway through -- the simulation.

While the gas is present, it damps planetary eccentricities and inclinations according to $\dot{e}/e=-1/\tau$ and $\dot{i}/i=-2/\tau$, where the damping timescale is
 \[
     \tau=0.003d\Bigg(\frac{a}{\rm au}\Bigg)^2\Bigg(\frac{M_{\odot}}{M_p}\Bigg) {\rm yr} \times   
     \begin{cases} 
      1 & v\leq c_s, \\
      (v/c_s)^3 & v> c_s, i < c_s/v_K,\\
      (v/c_s)^4 & i > c_s/v_k,
   \end{cases}
 \]
\noindent and $M_{\odot}$ is the mass of the Sun, $M_P$ is the mass of the planet, the Keplerian velocity $v_K=na$ where n is the planet's mean motion, the random epicyclic velocity $v = \sqrt{e^2+i^2} v_K$ where $i$ is the inclination, and $c_s~=~1.29$km/s~($a$/au)$^{-1/4}$ is the gas sound speed \citep{Papaloizou2000,Kominami2002,Ford2007,Rein2012}. The gas disk can also cause migration but we have confirmed through some trial simulations that include migration that the slow migration expected in a depleted gas disk has a negligible effect on the final planet properties. 

To address the question of whether a diversity in disk solid reservoirs can account for the diversity of super-Earth properties, we run ensembles of 80--400 simulations (detailed in Section \ref{subsec:sims}), each with a fixed value of $\alpha$ and $d$ and a continuum of $\sigz$ drawn randomly from a log uniform distribution. Using an initially log uniform distribution among simulations in an ensemble provides a set of simulations spanning a scale free range of $\sigz$. Later we will reweight the values of $\sigz$ to better match the properties of observed super-Earths (Section~\ref{subsec:compare}, i.e., to preferentially include simulations with certain $\sigz$ in our synthetically observed population). We list the  ensembles in Table~\ref{tab:simstuff}.

\begin{deluxetable}{lcccccc}
\tablecolumns{7}
\tablewidth{0pt}
\tablecaption{  \label{tab:simstuff}}
\tablehead{
\colhead{No.} &
\colhead{$e_0$} & 
\colhead{$i_0$} & 
\colhead{$\Delta_0$} & 
\colhead{$\sigz$} & 
\colhead{$\alpha$} &
\colhead{$d$} \\
\colhead{} &
\colhead{} &
\colhead{(rad)} &
\colhead{($R_H$)} & 
\colhead{g cm$^{-2}$} &  
\colhead{} &
\colhead{}
} 
\startdata
\lesserone & 0 & $0.01h/\sqrt{3}$ & 3   & 14--284 & -1.5 & 100  \\
\lessmany     & 0 & $0.01h/\sqrt{3}$ & 3 & 14--284 & -1.5 & 10 \\
\lessertwo     & 0 & $0.01h/\sqrt{3}$ & 3 & 14--284 & -1.5 & 1000 \\
\lesseroneslope & 0 & $0.01 h/\sqrt{3}$ & 3   & 2--43 & -2.5   & 100\\ 
\ngdtwofive     & $\sqrt{\frac{2}{3}}h$ & $h/\sqrt{3}$  & 10   & 12--54 & -2.5 & $^*$ 
\enddata
\tablecomments{The planetary embryos' initial eccentricities $e_0$ and inclinations $i_0$ are in terms of $h = \Big(\frac{M_{p,1}+M_{p,2}}{3M_{\star}}\Big)^{1/3}$ and spacing $\Delta_0$  is in terms of mutual Hill radii $R_H$ (Eqn. \ref{eq:space}). The disk parameters are the solid surface density normalization ($\sigz$) and radial slope $\alpha$ (Eqn. \ref{eq:alpha}) and gas depletion $d$ during the gas disk stage (Eqn. \ref{eq:gas}). \\
$^*$ Does not include gas damping.}
\end{deluxetable}

\subsection{Simulations of late stage planet formation}
\label{subsec:sims}

We simulate growth from isolation mass embryos to fully-fledged planets using {\tt mercury6} with the hybrid symplectic integrator \citep{Chambers1996}. Our simulations begin with embryos embedded in a gas disk and spaced by  
\begin{equation}\label{eq:space}
    \Delta \equiv \frac{a_2 - a_1} {R_H},
\end{equation}
\noindent where the mutual Hill radius is
\begin{equation}\label{eq:rh}
    R_H \equiv \frac{a_1+a_2}{2} \Big(\frac{M_{p,1}+M_{p,2}}{3M_{\star}}\Big)^{1/3},
\end{equation} 
and $M_\star$ is the stellar mass.  Based on the disk properties $\sigz$ and $\alpha$ (Section~\ref{subsec:disk}), the mass of each embryo is
\begin{eqnarray}
\label{eqn:memb}
M_p =
0.16 M_{\oplus} \left(\frac{\Delta_0}{3}\right)^{3/2}\left(\frac{\sigz}{33\, {\rm g}/{\rm cm}^2}\right)^{3/2} \nonumber \\
\times \left( \frac{a}{\rm au}\right)^{\frac{3}{2}(2-\alpha)} \left( \frac{M_\star}{M_\odot} \right)^{-1/2} .
\end{eqnarray} The semi-major axis of the innermost embryo is randomly drawn from a uniform distribution spanning 0.04--0.06 au, and the semi-major axis of each subsequent embryo is calculated according to Eqn.~\ref{eqn:memb} and $\Delta_0$ out to 1 AU. The resulting number of embryos ranges from 70 to 370. The embryos are initially spaced by $\Delta_0=3$ unless otherwise stated (see Table~\ref{tab:simstuff}). 

We start with such tight spacings -- much tighter than a typical isolation mass of $\Delta_0 \sim 10$ (e.g., \citealt{Kokubo1998})-- to allow the proto-planets to reach their own self-consistent isolation mass (with self-consistent eccentricities and inclinations) in the presence of gas. The initially tight spacings is an initialization step and is based on detailed exploration of initial spacings in DLC16 Sections 4.3 and 5.2. The embryos merge quickly, reaching a typical $\Delta=8$ within $\sim0.03$ Myr. This quick initialization leads to embryos with spacings, eccentricities, and inclinations dictated by the gas depletion level for subsequent evolution and represents an approximation to the more realistic scenario where proto-planets' $\Delta$ evolves as the gas surface density declines according to a photoevaporation model. As long as $\Delta_0$ is not artificially large, the final results are not sensitive to $\Delta_0$ in the gas stage; the gas stage evolution self-consistently dictates the initial conditions of the post-gas stage (DLC16, 5.2).

We integrate for 28 Myr with a timestep of 0.5 days and a close-encounter distance of $1 R_H,$ which switches the integrator from the symplectic integrator to the Burlicsh-Stoer integrator. We tested previously that using the Burlicsh-Stoer integrator for the entire simulation yields the same results (DLC16). Because of the short orbital periods simulated, we found previously that integrating for 10 times longer had a negligible impact on the final distribution of planet properties (DLC16). During the 1 Myr gas disk stage, we impose $\dot{e}$ and $\dot{i}$  (Section \ref{subsec:disk}) with \citet{Wolff2012}'s user-defined force routine. We assume perfect accretion without fragmentation when two embryos touch, using the {\tt mercury6} default density of 1 gcm$^{-2}$ in computing the collisional radius.

Following \citet{Dawson2015} and DLC16, we generate planet radii during post-processing using \cite{lee14}'s accretion models. We assume solar metallicity opacities with dust, a nebular temperature of 1500 K within 0.1 AU and 1500 K $\sqrt{0.1{\rm AU}/a}$ beyond 0.1 AU, and a nebular gas density according to Eqn.~\ref{eq:gas}. We neglect the core luminosity (as justified in \citealt{lee14}).

We find that after mergers, the new, more massive core quickly regains any atmosphere that would have been lost during the collision. Our ensemble results are insensitive to the treatment of atmospheric loss during gas stage collisions. For post gas stage collisions, we compare three treatments. In our nominal treatment, we assume that when two cores merge, the new core maintains the atmospheric mass of the larger core. This treatment assumes that not much gas is lost and/or that gas that escapes can be re-accreted as the planet orbits. In the second treatment, the new core retains half the atmospheric mass of the larger core. In the third treatment, the new core is totally stripped of its atmosphere if any collision(s) occur in the post gas stage. We discuss which results are sensitive to the treatment of post gas stage collisions in Section~\ref{sec:compositions}.

\subsection{Comparison to \kep sample}
\label{subsec:compare}

When we assess how the conditions for late-stage planet formation manifest in the \kep observables (Sections \ref{sec:orbits} and \ref{sec:compositions}), we will make comparisons to the DR25 \kep catalog \citep{thom18}. We limit the \kep catalog to systems with stellar effective temperature $4100K < T_{eff}<6100K$, stellar $logg >4$, Kepler magnitude $<$ 15, and containing at least one planet with $R<4R_{\oplus}$. Statistical modeling has shown that less than 10\% of Kepler's planetary candidates are false positives \citep{Morton2011, Fressin2013}, allowing us to represent the true planet population with this sample.

We convert the simulated planets into transiting planets using the following forward modeling procedure. First, we generate $10^4$ systems randomly oriented in space for each simulated system; planets with impact parameters b $<1$ are said to be transiting. We compared this method to \citet{Brakensiek2016}'s CORBITS and found that it produced consistent results.  

Next we assess which planets have sufficient signal-to-noise to be detected. DLC16 applied simple detection cuts of period P~$<200d$ and planetary mass $2M_{\oplus}<M_P<30M_{\oplus}$. Instead of these simple cuts, we compute the probability of detection for each planet following \citet{Burke2015}, who model the probability of detection using the \citet{Christiansen2015,chri16} completeness parametrization; we use the appropriate completeness function parameters for DR25. (We do not model the reduced detection efficiency of subsequent planets in the same system \citep{zink19} and discuss the effect on our results in Section \ref{subsec:orbresults}.) See Appendix for an empirical exploration of detection efficiency.

\begin{figure}
    \centering
    \includegraphics[width=0.48\textwidth]{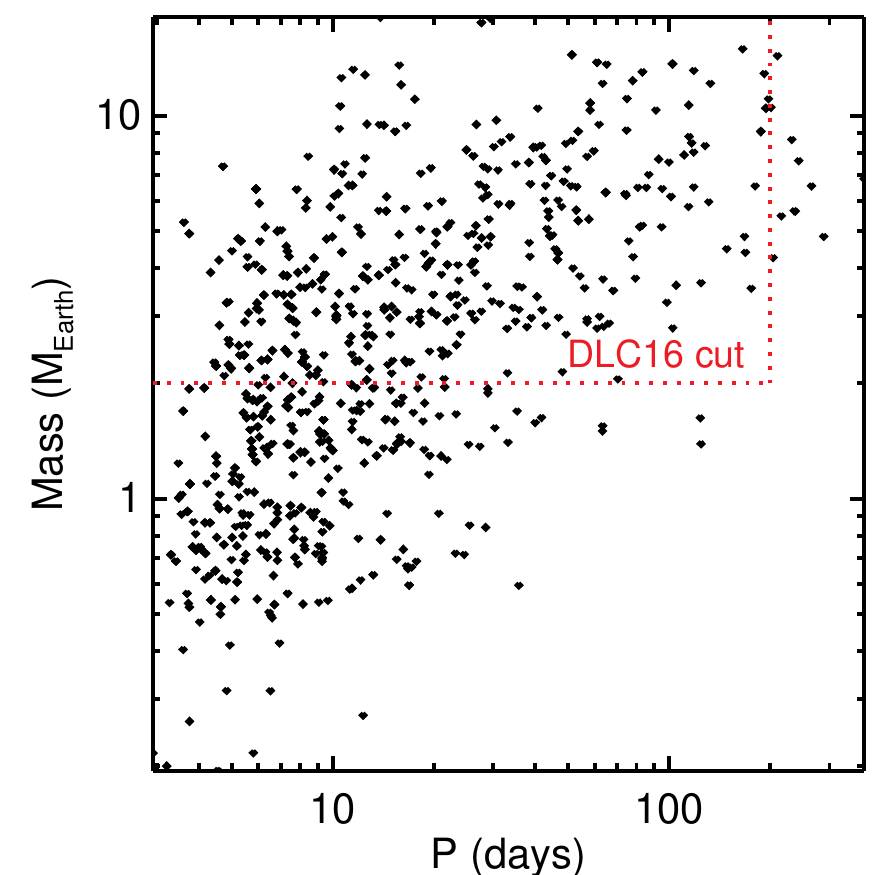}
    \caption{Mass vs. orbital period of simulated detected transiting planets in Ensemble \lesserone. The red dotted lines show DLC16's detection cuts of $2M_{\oplus}<M_p$ and orbital period $P<200$ days. By modeling the detection efficiency instead of applying simple cuts, we include more low mass planets at short orbital periods.}
    \label{fig:kports}
\end{figure}

We compute the detection probability on a grid of impact parameter, planet radius, and orbital period, marginalizing over the combined differential photometric precision values for all \kep target stars with $4100K < T_{eff}<6100K$, stellar $log g > 4$, and Kepler magnitude $<$ 15. We assign each simulated planet a radius using the \citet{Weiss2014} mass-radius relationship \footnote{Because our simulations do not include photo-evaporation, our simulated radii from the \citet{lee14} are not realistic for short period planets. Using a mass-radius relationship allows to compare to the short period planets that make up the bulk of the \kep sample.}. For each of the $10^4$ realizations that results in one or more transiting planets, we draw a random uniform number between 0 and 1 and detect those transiting planets with detection probability greater than that number.  We compare our distribution of simulated planets using DLC16's selection cuts vs. our more realistic detection efficiency in Fig. \ref{fig:kports}. The more realistic detection efficiency allows us to include the subset of low-mass and/or long period planets that would be detectable around a subset of \kep target stars. 

\section{Establishing the Diversity of Super-Earth Orbital Properties} \label{sec:orbits}

Here we use the disk properties, simulations, and the planet candidate catalog described in Section~\ref{sec:simulations} to investigate whether a continuum of planet formation conditions can account for the observed diversity of \kep super-Earth orbital properties. As motivated in Section~\ref{sec:intro}, we use a range of the solid surface density normalizations to allow us to simulate a continuum of formation conditions: from low $\sigz$ Solar System like conditions where late stage formation via giant impacts is delayed until the post-gas stage, to high $\sigz$ conditions where compact super-Earths finish forming in residual disk gas.

In Section~\ref{subsec:obs}, we describe the key observable orbital properties we use for comparison. In Section~\ref{subsec:reweight}, we describe how we adjust the underlying distribution of disk conditions to match the observables. In Section~\ref{subsec:orbresults}, we show that a continuum of formation conditions parametrized as $\sigz$ can match the assortment of observed orbital properties. We discuss the influence of other disk properties in Section~\ref{subsec:other}. We show that a continuum in formation conditions can lead to distinctions between eccentricities  of single vs. multi transiting systems in Section~\ref{subsec:ei}. 

\subsection{Key \kep Observables}
\label{subsec:obs}

We compare our simulated transiting planets to orbital properties of the \kep catalog (Section~\ref{subsec:compare}) using several key population-wide observables: transit multiplicity, ratio of transit duration between adjacent planets, period ratio of adjacent planets, and Hill spacing between adjacent planets.

\begin{figure*}
    \centering
    \includegraphics{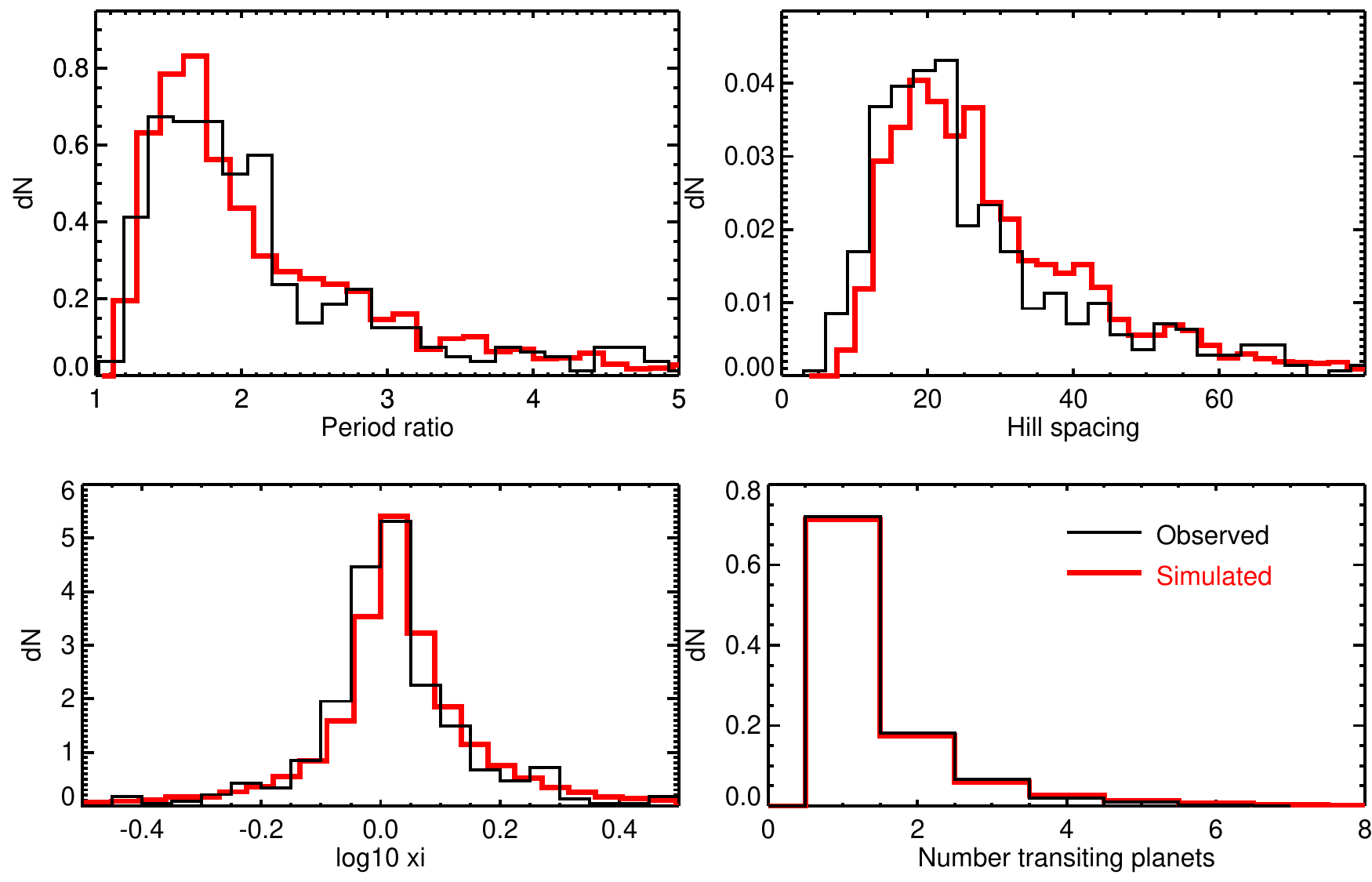}
    \caption{Period ratios, spacing in mutual Hill radii, transit duration ratio ($\xi$), and transit multiplicity observed (black) and simulated from ensemble \lesserone (red). The simulated distributions match the observed, demonstrating that a continuum of formation conditions can account for the observed diversity of orbital properties.}
    \label{fig:hists}
\end{figure*}

A system's transit multiplicity is the number of transiting planets per system. The multiplicity distribution of \kep planet candidates peaks sharply for one transiting planet (singles), but contains a tail of multi-transiting systems (Figure~\ref{fig:hists}, bottom right panel). This peak of singles drives many studies of the \kep dichotomy (e.g., \citealt{Pu2015,Volk2015,ball16},MB16,\citealt{Spalding2016,Huang2017}) because single component parametric models of underlying planet multiplicity and single mode formation models often produce too few singles (e.g.,\citealt{Lissauer2011,Fang2012,joha12,Hansen2013,he19}; see \citealt{zink19} for a single component model that matches the observed multiplicity but does not attempt to match the duration and period ratios). The transit multiplicity is set by the underlying planet multiplicity and the planets' mutual inclinations. 

Another observed property affected by planets' mutual inclinations is the normalized transit duration ratio $\xi$ (e.g., \citealt{Fang2012,Fabrycky2014}), defined as
 
\begin{equation}\label{eq:tdur}
     \xi =  \frac{T_{dur,1}}{T_{dur,2}} \left(\frac{P_1}{P_2}\right)^{-1/3}
\end{equation}
 
\noindent where $T_{dur}$ is the transit duration and $P$ is the orbital period and 1 and 2 note the inner and outer planet respectively. The observed distribution $\log \xi$ peaks around 0 and is asymmetric (Fig. \ref{fig:hists}, bottom left). Coplanarity shifts the distribution to $\xi \ge 1$, and eccentricities widen the distribution. Because observational uncertainties also widen the $\xi$ distribution, we add mock observational uncertainties to our simulated transit duration ratios. For each planet pair in the observed sample (defined in Section \ref{subsec:obs}), we compute the uncertainty in the transit duration ratio from each planet's reported duration, upper uncertainty, and lower uncertainty \citep{thom18}. We compile a distribution of transit duration ratio upper uncertainties $\sigma_{\xi, obs, +}$ and lower uncertainties $\sigma_{\xi, obs, -}$. For each simulated pair with duration $\xi$, we draw a transit duration uncertainty $\sigma_{\xi, obs}$ and draw the transit duration from an asymmetric normal distribution with mean $\xi$ and standard deviation $\sigma_{\xi, obs,+}$ above the mean and $\sigma_{\xi, obs,-}$ below the mean.

The distribution of period ratios of observed adjacent \kep planets peaks at around 1.7 (Fig. \ref{fig:hists}, top left panel), with a tail that extends to much larger values. Although orbital resonances likely shape the finer structure of distribution near 3:2 and 2:1 (e.g., \citealt{Fabrycky2014}), here we focus on reproducing the broader distribution. Previous studies have found it challenging to match the peak (e.g., \citealt{Hansen2013}) and to produce the most compact systems with ratios $<1.3$ (e.g., D16).

The period ratio directly relates to the Hill spacing (Eqn. \ref{eq:space}) through the planets-to-star mass ratio (e.g., \citealt{Malhotra2015}). The Hill spacing is typically the more determinate quantity for orbital evolution, except near mean motion orbital resonances and for the tightest period ratios where mean motion resonance overlap affects stability (e.g.,\citealt{Deck2015}). The timescale for orbit crossing increases exponentially with the Hill spacing (e.g., \citealt{Chambers1996,Yoshinaga1999,Zhou2007,Pu2015}), and therefore the spacing must be wide enough to persist over the system's age. DLC16 argued that although instabilities and mergers can occur throughout the system's lifetime\footnote{DLC16 did not simulate for a full several Gyr lifetime due computational limitations, but found little evolution in the distributions of observables from 10 Myr to 300 Myr.}, the distribution of Hill spacings is primarily locked in during the formation stage. The spacing is dictated by an eccentricity equilibrium that balances gravitational scatterings that excite\footnote{ Other physical processes that excite eccentricities, such as secular perturbations from distant giant planets, can contribute the eccentricity equilibrium as well.} eccentricities with mergers that damp eccentricities. When formation takes place with residual disk gas still present, planets can achieve a tighter spacing that persists over the system's lifetime. Hence planet formation conditions dictate the final spacings.

To compute the Hill spacing for the observed \kep planets, we estimate the planet mass from the planet radius using the \citet{Weiss2014} mass-radius relation; the resulting distribution is insensitive to the assumed mass-radius relationship (e.g., \citealt{Pu2015}). For the \kep population, the distribution of Hill spacing is right-skewed and peaks around 15--20~$R_H$ (e.g., \citealt{fang13}), smaller than our Solar System terrestrial planets' average spacing of 43~$R_H$. The observed distribution has a tail that extends to large values beyond 100, exceeding Mercury and Venus' Hill spacing of 63. Simulations of planet formation without disk gas lead to spacings more like Solar System's \citep{Hansen2013}, but including gas allows for the more compact spacings observed (DLC16).

Previously, DLC16 combined simulated transiting planets from two types of formation conditions to produce observables mostly consistent with those observed (see also MB16), though they were not able to reproduce the tightest period ratios. Here we will revisit these observables using a wider, continuous, and more flexible distribution of formation conditions.

\subsection{Generating a flexible distribution of formation conditions with reweighting} 
\label{subsec:reweight}

As motivated in Sections \ref{sec:intro} and \ref{subsec:disk} and detailed in Section \ref{sec:simulations}, we run ensembles of simulations. Each simulation within an ensemble has a different value of $\sigz$, the solid surface density normalization -- with each ensemble spanning a log uniform range -- but the same value of the radial profile of solids $\alpha$ and the gas depletion $d$ (Table \ref{tab:simstuff}). Forming planetary systems ranging from the Solar System to systems of close-in, compact super-Earths requires a large variation in $\sigz$ among stars, but the true underlying distribution is unknown to us. 

We reweighted the simulations' underlying distribution of $\sigz$ using the function
\begin{equation}\label{eq:ssd}
    W = \exp\left[{-\frac{(\sigz-\bar{\sigz})^2}{2 \sigma_{\sigz}^2}}\right]
\end{equation}
\noindent where $\bar{\sigz}$ and $\sigma_{\sigz}$ are constants. We use this weighting function to preferentially include simulations with $\sigz$ close to $\bar{\sigz}$ in our population of synthetic transiting planets. Each simulated system has a different value of $\sigz$. Recall that in Section \ref{subsec:compare}, we generated $10^4$ randomly oriented realizations of each simulated system. For each realization in which at least one planet transits and is detected, we draw a random uniform number between 0 and 1. We include the realization if the random number is less than $W$.

We adjusted $\bar{\sigz}$ and $\sigma_{\sigz}$ by hand to produce a qualitatively good match to the four distribution of observables. We find that $\bar{\sigz} = 0 $ and $\sigma_{\sigz} = 65$ g cm$^{-2}$ enable Ensemble \lesserone to match the observables satisfactorily. Even with the reweighting, other ensembles exhibit discrepancies with the observations that we will explore further in Section \ref{subsec:other}.

\subsection{A diversity of orbits from a continuum of formation conditions} \label{subsec:orbresults}

\begin{figure}
    \centering
    \includegraphics[width=0.49\textwidth]{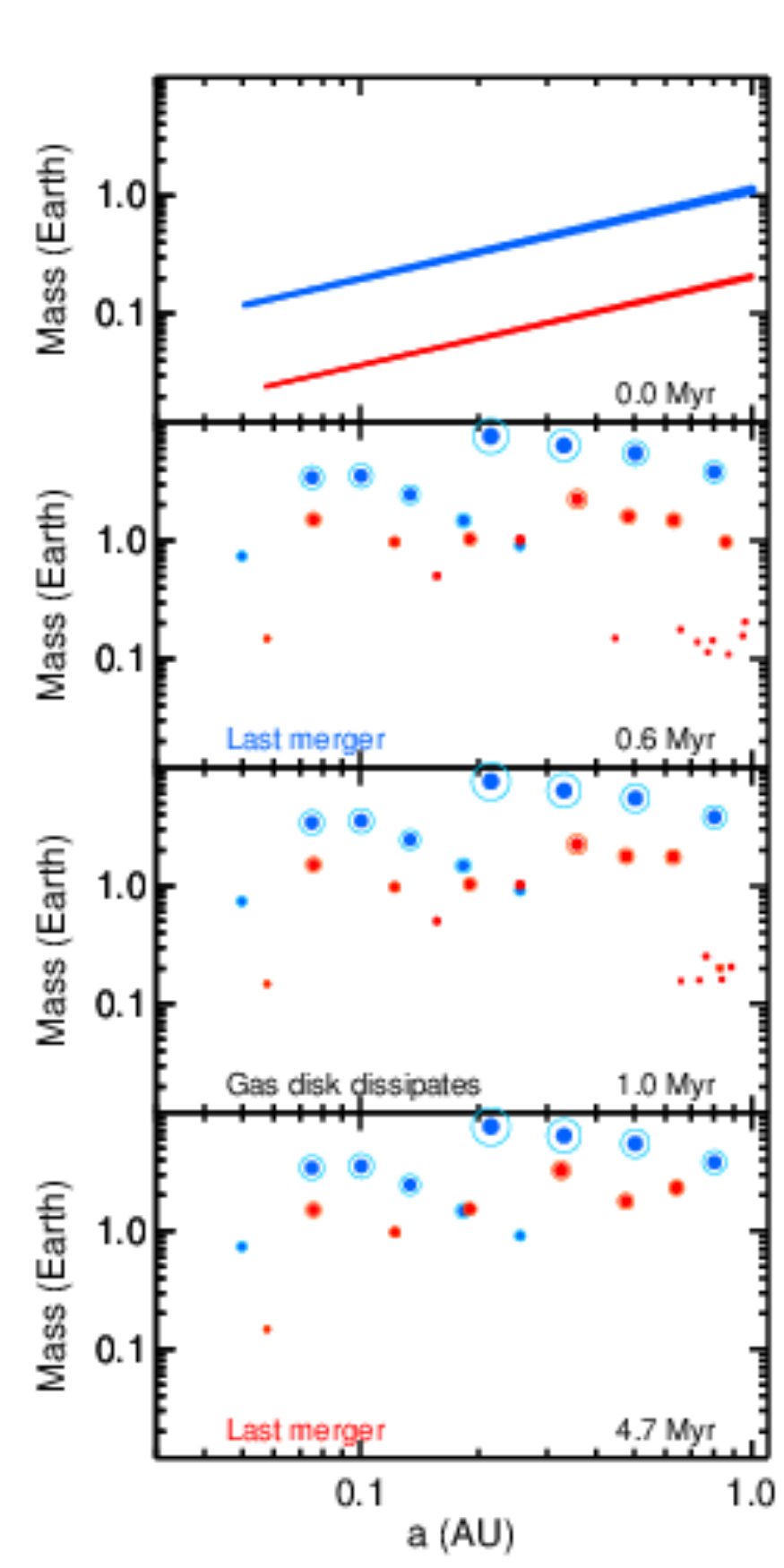}
    \caption{Snapshots of planet mass vs.~semi-major axis from two simulations, color-coded by solid surface density normalization $\Sigma_{z,1}$:  121 gcm$^{-2}$ (blue) and 39 gcm$^{-2}$ (red). In the blue simulation, planets finish forming during the gas disk stage and end up on dynamically cold (tightly spaced, low eccentricity, low mutual inclination) orbits with gas envelopes. In the blue simulation, planets finish forming during the gas disk stage and end up on more tightly spaced orbits, circular ($<e> = 0.004$ vs $<e> = 0.009$ for the red simulation) with more gas. The symbol size is proportional to the planet core radius (inner) and total radius (outer).}
    \label{fig:ex}
\end{figure}

\begin{figure*}
    \centering
    \includegraphics[width=0.99\textwidth]{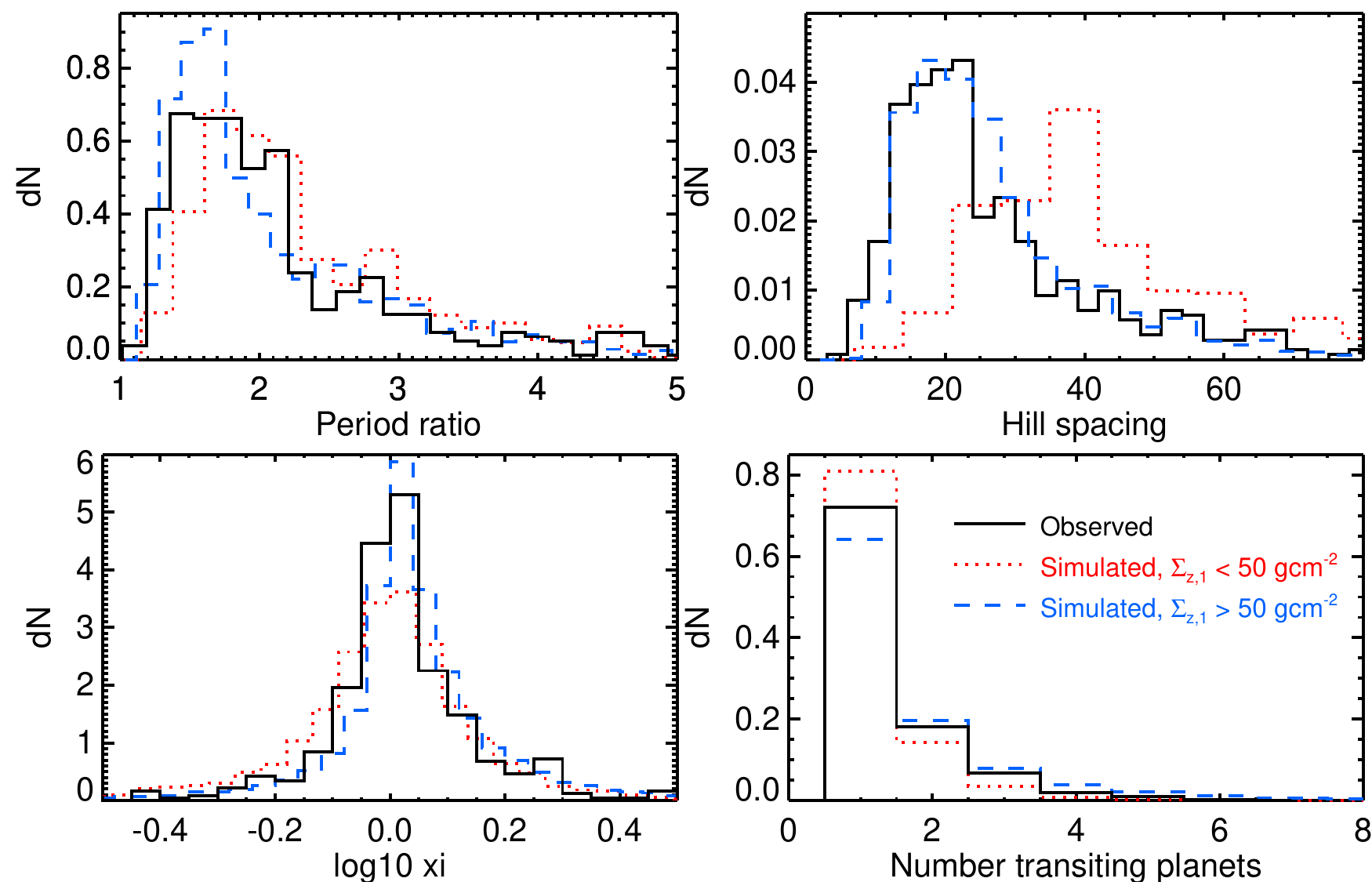}
    \caption{More material in the region of in situ formation lead to more tightly-spaced, coplanar, high transit multiplicity systems. Period ratios, spacing in mutual Hill radii, transit duration ratio ($\xi$), and multiplicity of resulting simulations from Ensemble \lesserone for large solid surface density normalization ($\sigz>50$ g/cm$^2$) in blue dashed and for small solid surface density normalization ($\sigz<50$ g/cm$^2$) in red dotted where 50 g/cm$^2$ marks the average solid surface density normalization among transiting systems. The systems from the catalog are shown in black.}
    \label{fig:hists_split}
\end{figure*}

We find that a continuum of values in $\sigz$, i.e., disks ranging from small to large solid reservoirs in the region of in situ formation, can account for the observed diversity in orbital properties (Figure~\ref{fig:hists_split}). Our reweighted distribution of $\sigz$ (Section \ref{subsec:reweight}) corresponds to an underlying distribution that peaks at small values similar to the that assumed for our solar system ($\sigz = 10$ gcm$^{-2}$, e.g., \citealt{chian10}) but extends to larger values. Although intrinsically rarer, the larger $\sigz$ disks produce more detected transiting planets (because the planets tend to be larger); the median among simulated systems of transiting planets is $\sigz$ =50 gcm$^{-2}$. Figure \ref{fig:ex} shows snapshots of the evolution of two example systems.

We plot the observables split by that median $\sigz$ among simulated systems of transiting planets. We find that planets formed in local regions richer in solids ($\sigz>50$ g/cm$^2$) tend to be dynamically cold: they have smaller Hill spacings, an asymmetric distribution of duration ratios (caused by their low mutual inclinations and eccentricities), and high transiting planet multiplicity. Planets formed in regions poorer in solids ($\sigz<50$ g/cm$^2$) tend to be dynamically hot: they have wider Hill spacings, a wider and more symmetric distribution of duration ratios (caused by their larger mutual inclinations and eccentricities), and more single transiting planets. (Note that because higher $\sigz$ leads to high transiting planet multiplicity, the median simulated solid surface among synthetically observed transiting planet pairs is 82 g/cm$^2$. Therefore there are more pairs represented in the blue histogram than the red histograms of period ratio, duration ratio, and Hill spacing.) The peak in the distribution of period ratios is similar for low vs. high $\sigz$ because $\sigz>50$ g/cm$^2$ systems tend to have larger Hill spacings but also larger planet masses.  However, the tail of the period ratio distribution extends to larger values for $\sigz<50$ g/cm$^2$ systems. Higher solid surface densities allow planets to complete their formation via giant impacts in the presence of residual disk gas. The amount of gas around is high enough to aid the planets in reaching compact, circular, coplanar configurations but not so high that it cushions them against mergers (i.e., timescale for eccentricity damping is too long to prevent mergers). 

Unlike in DLC16, our simulations produce the tightest observed period ratios $<1.3$. As we would expect from stability, the smaller period ratio planets tend to have lower masses. Thus implementing a more realistic detection efficiency than DLC16's 2 $M_\oplus$ cut yields more simulated transiting planets with period ratios $<1.3$. We also find that the tightest period ratios occur at an intermediate range of $30 < \sigz < 100$ gcm$^{-2}$: lower $\sigz$ systems are too dynamically hot and higher $\sigz$ systems are too massive for very tight period ratios. Extending our range of $\sigz$ (14--284 gcm$^{-2}$ vs. DLC16's 38--105 gcm$^{-2}$) also produced additional tight period ratio systems.

Compared to the observed distribution, we find a shortage of planets with Hill spacings of $\Delta < 15$. A two sample  Kolmogorov-Smirnov rejects the null hypothesis that the observed and simulated samples are drawn from the same underlying distribution with $p=10^{-6}$ (but cannot reject the null hypothesis for $\Delta > 15$). The Hill spacing is not a directly observed quantity but computed from the observed period ratio and masses with an assumed mass-radius relationship. The agreement of the observed and simulated period ratios but disagreement of the Hill spacing implies that our tightly spaced simulated planets have lower masses than those assumed for the observed planets. In Section \ref{subsec:mr}, we will show that our small period ratio simulated planets do have lower masses at a given radius and that there is tentative evidence that observed planets may exhibit the same trend \citep{Weiss2014,mill17}. Therefore the disagreement may lie in the assumed mass-radius relationship for observed planets (Fig. \ref{fig:mr}). Alternatively, we may be able to produce tight period ratios at large masses if we were to fine tune our simulation parameters; in Section \ref{subsec:other}, we will show that gas depletion factor $d$ affects the mass range of planets in compact systems. A final possibility is that in situ formation produces a true deficit of small $\Delta$ systems and that observed systems have a different origin, such as migration.

\citet{zink19} found that accounting for the reduced detection efficiency of subsequent planets could reduce the ratio of single-to-double transiting planets by 10\% and that the effect was largest for planets with periods $>200$ days. We do not model this effect. To check the robustness of our results, we compare only simulated and observed planets with periods $<200$ days and find no noticeable effect on our ensemble results. We also find that we can match a 10\% reduction in singles by slightly adjusting the weightings we use in Section \ref{subsec:reweight}.

\subsection{Influence of other disk properties}
\label{subsec:other}

We have chosen to generate a continuum of planet formation conditions using a continuum of disk solid surface density normalizations (as justified in Section \ref{sec:intro}), keeping other disk properties (Section \ref{subsec:disk}) fixed for each ensemble of simulations. Even with reweighting of the solid surface density normalizations ($\sigz$), only Ensemble \lesserone (gas depletion $d=100$, solid surface density slope $\alpha=-1.5$) provides a satisfactory qualitative agreement with the observations. Here we explore the effects of other disk parameters on the observables.

Ensembles with less depleted (Ensemble \lessmany) or more depleted (Ensemble \lessertwo) gas disks fail to match the observables. Dynamically cold systems require solid-to-gas ratio $\sim$3--15: high enough so that planets can complete their assembly by collisional mergers yet low enough so that nebular gas can damp away planetary eccentricities and mutual inclinations. Acquiring this solid-to-gas ratio is necessary but {\emph not} sufficient to produce the observed orbital properties of {\it Kepler} planets. In less depleted disks like Ensemble \lessmany, high solid surface density disks spawn dynamically cold systems of massive planets; because of the large masses, the tightest observed period ratio systems are not reproduced. In more depleted disks like Ensemble \lessertwo, low solid surface density disks produce dynamically cold systems of low mass planets. For a given Hill spacing, these systems contain more planets, and for a given Hill velocity, the mutual inclinations are lower. Matching the observed distribution of period ratios therefore results in too few singles and an insufficient spread in transit duration ratios. 

Using a different form for reweighting than Eqn. \ref{eq:ssd} would not solve the discrepancies between these ensembles and observed \kep sample. For less gas-depleted conditions (Ensemble \lessmany), no value of $\sigz$ produces the tightest period ratios. For more gas-depleted conditions (Ensemble \lessertwo), only low mass systems (resulting from low $\sigz$ disks) can achieve tight period ratios, resulting in mutual inclinations that are too small; adding more high $\sigz$ systems to the mix can widen the distribution of duration ratios, but skews the distribution of period ratios to inconsistently large values.

DLC16 combined two ensembles each with the same range of $\sigz$ but different gas depletion factors to account for the full range of observables, whereas our Ensemble \lesserone accounts for the observables using a wider and more flexible underlying distribution of $\sigz$. One of their two ensembles -- with a gas depletion factor $d$ corresponding to a solid-to-gas ratio $\sim$2--6 for their range of solid surface densities -- primarily produced dynamically cold systems, and the other -- with a gas depletion factor $d$ corresponding to a solid-to-gas ratio $\sim$200--600 for the same range of solid surface densities - primarily produced dynamically hot systems. Thus both we and DLC16 account for the observables via formation in environments with a diversity of solid-to-gas ratios but -- for reasons justified in Sections \ref{sec:intro} and \ref{subsec:disk} -- we choose to produce the variation through the amount solids instead of the amount of gas.

We also explore the effect of the steeper $\alpha=-2.5$ solid surface density profile introduced in MB16, who achieved a mixture of dynamically hot and cold systems with a mixture of two profiles. In Ensemble \ngdtwofive, we follow MB16 and set the initial Hill spacing $\Delta_0$=10, equivalent to picking up the simulation after a gas disk stage that allows for growth but keeps eccentricities and inclinations very low (DLC16). In Ensemble \lesseroneslope, we simulate the gas disk stage; Ensemble \lesseroneslope is equivalent to Ensemble \lesserone but with $\alpha=-2.5$. For both ensembles, we find that the weighting of solid surface densities toward the low values that are necessary to match the observed period and duration ratios results in too few singles. The steep surface density profile distributes many planets close in, resulting in too many multi-transiting systems.

Because we have not exhaustively explored the full parameter space of disk properties, we cannot conclude that Ensemble \lesserone's disk properties are the \emph{only} possibility for matching the \kep observables in our general model. In the future, we will explore a wider parameter space to infer the underlying distribution of disk properties. In Section \ref{sec:conclusion}, we lay out a plan for conducting such an exploration given limited computational resources.

\subsection{Eccentricities of singles vs. multis}
\label{subsec:ei}

\begin{figure}
    \centering
    \includegraphics[width=0.49\textwidth]{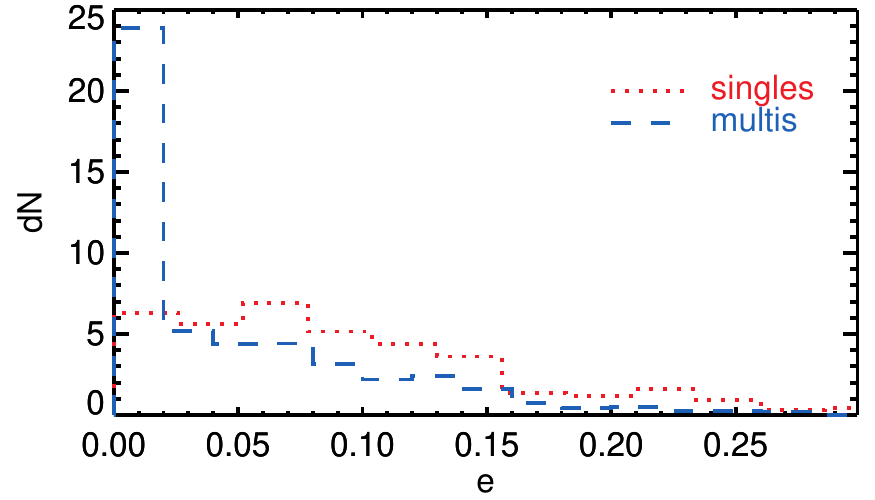}
        \includegraphics[width=0.49\textwidth]{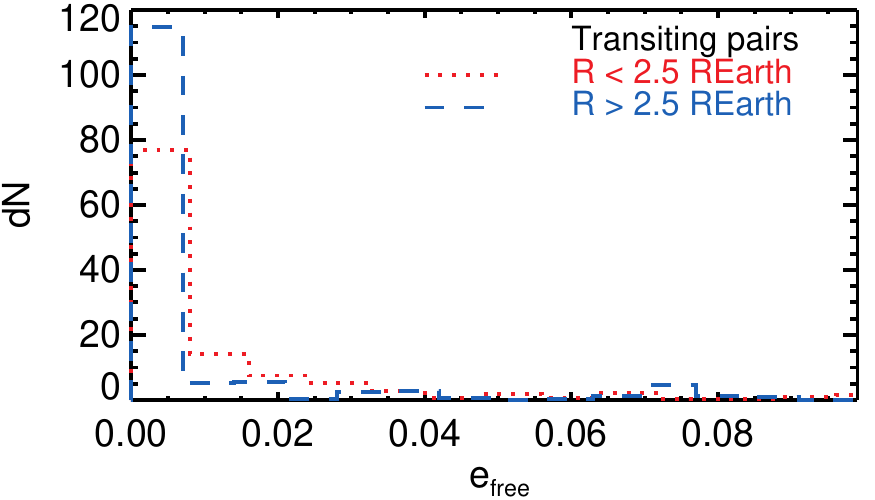}
    \caption{Top: Eccentricities of simulated planets split by multi-transiting ("multis") in dashed blue and tranets in single-transiting systems ("singles") in dotted red. Singles have a wider ranges of eccentricities than multis. Bottom: Eccentricities split by planet size for transiting planet pairs. We find that smaller planets tend to have larger eccentricities. Larger planets form early on and are therefore subject to damping by the residual gas disk, leading to the smaller eccentricities.}
    \label{fig:ecc}
\end{figure}

In \citet{vaneylen2019}, we showed that in situ formation can account for the observed trend that multi-transiting systems tend to have lower eccentricities. Using our updated Ensemble \lesserone (with improved detection efficiency modeling and reweighting of solid surface densitites), we plot the eccentricity distributions in Fig. \ref{fig:ecc} and find that this trend holds. The disks that form dynamically cold systems have both lower eccentricities and lower mutual inclinations and therefore tend to manifest as multi-transiting. Dynamically hot systems tend to have higher eccentricities and higher mutual inclinations; they tend to manifest as singles. We find a mean $<e> = 0.05$ for multi-transiting systems and $<e> = 0.10$ for single-transiting systems. With respect to their system's initial plane, a mean $<i> = 2.3^\circ$ for multi-transiting systems and $<i> = 5.5^\circ$ for single-transiting systems. For comparison, our solar system's terrestrial planets have eccentricities ranging from 0.007 (Venus) to 0.21 (Mercury) and mutual inclinations ranging from 1.85$^\circ$ (Earth and Mars) to 7.0$^\circ$ degrees (Earth and Mercury), falling within the simulated distributions.

Our results for multi-transiting system are consistent with trends among observed Kepler planets, with distribution parameters corresponding to mean values of $<e> = 0.076 ^{+0.013}_{-0.025}$ \citep{vaneylen2019}, $<e> = 0.044 \pm 0.015$ \citep{mills19}, and $<e> =  0.04 ^{+0.04}_{-0.03}$ \citep{xie16}. However, results for single-transiting system are significantly lower than among observed Kepler planets, with distribution parameters corresponding to mean values of $<e> = 0.30\pm0.05$ \citep{vaneylen2019}, $<e> = 0.209 ^{+0.016}_{-0.010}$ \citep{mills19}, and $<e> =  0.32 \pm 0.02$ \citep{xie16}. The fundamental source of the discrepancy is that observed eccentricity distribution of single planets has a tail extending to eccentricities $e>0.3$ (e.g., \citealt{mills19} Section 3) that self-stirring during in situ formation cannot reproduce (e.g., \citealt{vaneylen2019}, Figure 9) because eccentricities are limited to the ratio of the escape velocity from the surface of the planet to the Keplerian velocity (e.g., \citealt{Petrovich2014,Schlichting14}). As argued by \citet{vaneylen2019}, another explanation is needed for these observed, highly elliptical systems, such as perturbations from outer giant planets. See \citet{ball19} for predictions of links between multiplicity and planet properties for TESS.
 
\section{Establishing the Diversity of Super-Earth Compositional Properties}\label{sec:compositions}

In our simulations, variations in formation conditions lead to an assortment of compositions \citep{Dawson2015} and link orbital properties to compositional properties (DLC16). We now have an ensemble of simulated planets that is produced from a continuum (rather than dichotomy) of formation conditions and that matches the observed diversity of \kep orbital properties (Section \ref{sec:orbits}). Here we revisit compositional trends explored briefly in DLC16 and explore new trends motivated by \kep follow up observations. We investigate links between eccentricity and size in Section \ref{subsec:size}, scatter in the mass-radius relation in Section \ref{subsec:mr}, and size similarity within each system in Section \ref{subsec:peas}. 

\subsection{Eccentricity-size links}
\label{subsec:size}

In their analysis of planets characterized by transiting timing variations, \citet{Hadden2014} found lower free eccentricities for planets for larger planets (a Rayleigh distribution characterized by width $\sigma_e=0.008^{+0.003}_{-0.002}$ when both planets have $R_p > 2.5 R_\oplus$, where $R_\oplus$ is the Earth's radius, corresponding to mean $<e> = 0.010$) than smaller planets ($\sigma_e=0.017^{+0.009}_{-0.005}$ when both planets have $R_p < 2.5 R_\oplus$ corresponding to mean $<e> = 0.021$ ). Our simulated planets exhibit the same trend: planets that finish their formation during the gas stage exhibit both larger radii (due to accreting gas) and lower eccentricities (due to gas damping).

To make a quantitative comparison, we limit our simulated planets to those beyond 0.15 AU because our radius models do not include photoevaporation. About 1/5 of the \citet{Hadden2014} sample has an inner planet with $a>0.15$ AU. We estimate the free eccentricity as the magnitude of the separations between the eccentricity vectors. Among pairs of simulated transiting planets with period ratios less than 2.5, we find a mean $<e_{\rm free}> = 0.012$ when both planets have $R_p > 2.5 R_\oplus$ and $<e_{\rm free}> = 0.022$ when both planets have $R_p < 2.5 R_\oplus$. These mean values are similar to those inferred by \citet{Hadden2014}, but we note that our eccentricity distributions are more skewed, peak at small values, and long-tailed than \citet{Hadden2014}'s assumed Rayleigh distribution (Fig. \ref{fig:ecc}). The trends persist for all of our treatments of post-gas stage atmospheric loss via collisions (Section \ref{subsec:sims}). The most extreme alternative treatment (any post-gas collision completely removes the atmosphere) produces lower eccentricities because planets that undergo significant dynamical evolution after the gas stage (leading to larger eccentricities) end up smaller and less likely to be detected.

\begin{figure}
    \centering
        \includegraphics[width=0.47\textwidth]{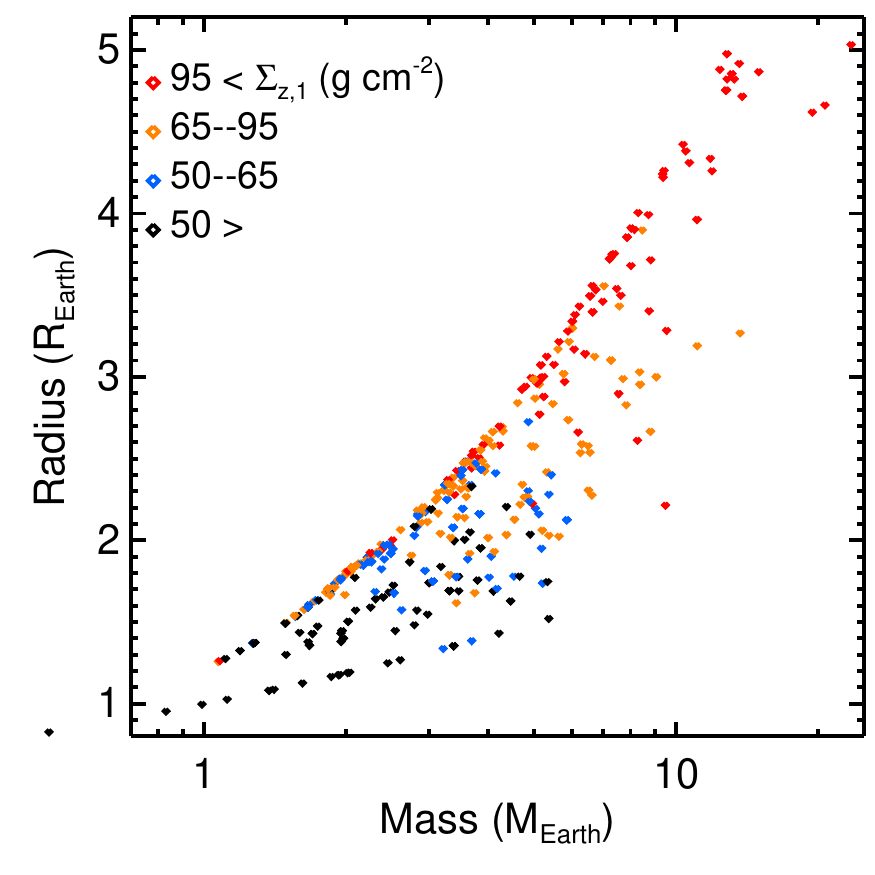}
    \includegraphics[width=0.47\textwidth]{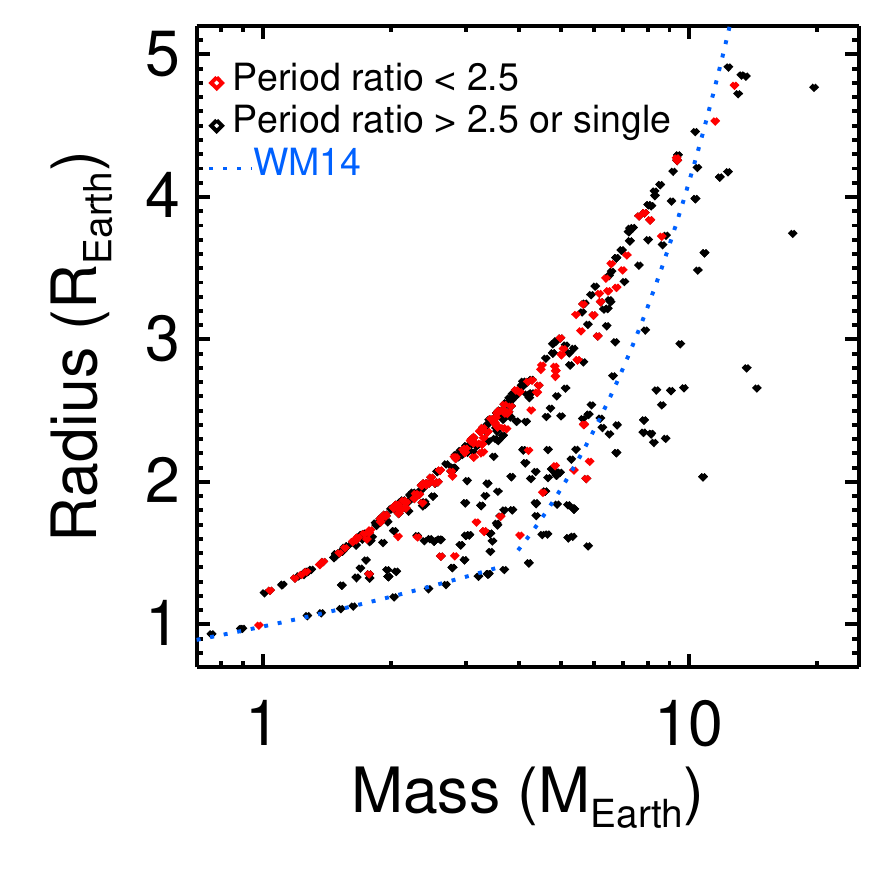}
     \caption{Mass-radius relationship of simulated planets from Ensemble \lesserone colored by the initial solid surface density normalization $\sigz$ (top) and period ratio (bottom). Top: The dispersion we see can be partially attributed to formation conditions, as planets that form in disks with larger $\sigz$ can more often acquire gas envelopes before the gas disk dissipates, leading to larger radii at a given mass. Bottom: Simulated planets in compact systems (red) have lower densities (i.e., larger radii at a given mass) than wider pairs or single transiting planets. The \citet{Weiss2014} used in computing the mutual Hill radii of observed planets in Section \ref{subsec:obs} is overplotted for comparison. The observed mass-radius relationship also exhibits intrinsic scatter \citep{Wolfgang2016}.
    \label{fig:mr}}
\end{figure}

\subsection{Scatter in the Mass-Radius Relation}
\label{subsec:mr}
\citet{Wolfgang2016} demonstrated that the observed mass-radius relationship for super-Earths contains intrinsic scatter (i.e., scatter beyond that introduced by observational uncertainties). 
In the in situ formation paradigm, planets of a given mass can form with significantly different radii depending on when the cores complete their assembly with respect to the time at which disk gas dissipate \citep[see also][their Figure 6]{Lee19}. Part of this variation in growth time is due to the stochastic nature of the giant impact stage, but part is due to variation in formation conditions established by disk properties (Section \ref{subsec:disk}). For cores that complete their assembly during the depleted gas stage, diversity in disks' gas dissipation timescales can further enhance this variation, an effect we recommend exploring in future studies.

We show simulated radii versus planet mass in Figure~\ref{fig:mr}; as in Section \ref{subsec:size}, our simulated planets are all beyond 0.15 AU. The mass-radius relation exhibits a wide dispersion that can be partially accredited to the amount of solids in the formation region where planets formed (parametrized as $\sigz$; Section \ref{subsec:disk}): at a given mass, planets that formed in disks with more solids tend to have larger radii (Fig.~\ref{fig:mr}, left panel). Embryos that form in disks with with more solids can reach their final masses in the presence of gas and acquire gas envelopes. 

Simulated transiting planets observed as pairs with period ratios less than 2.5 tend to have larger radii at a given mass than wider pairs or single transiting planets. This trend arises from the tendency of compact systems to finish forming during the gas disk stage. The compact systems are also more likely to exhibit transit timing variations. Therefore in situ formation in residual gas may contribute to the observed trend that planets with masses measured via transiting timing variations have lower densities \citep{Weiss2014,Wolfgang2016} than planets with masses measured via radial velocity. 

Post-impact atmospheric loss affects can have a big effect on individual planet radii \citep{inam15}. However, alternative treatments of post-impact atmospheric loss (Section \ref{subsec:sims}) all yield a similar spread in the mass-radius relationship. At a given mass, the radii range from purely rocky bodies to those with a full, undisturbed envelope accreted during the gas disk stage. The more extreme the atmospheric loss we assume, the fewer planets we see at intermediate radii. The range of radii can be accounted for by formation (i.e., whether or not the core grows large enough to accrete gas during the gas disk lifetime, e.g., \citealt{Dawson2015}) but the distribution within that range depends on efficiency of post-gas disk collisional atmospheric loss.

\subsection{Similarities in size and spacing}
\label{subsec:peas}

\begin{figure}
    \centering
        \includegraphics[width=0.47\textwidth]{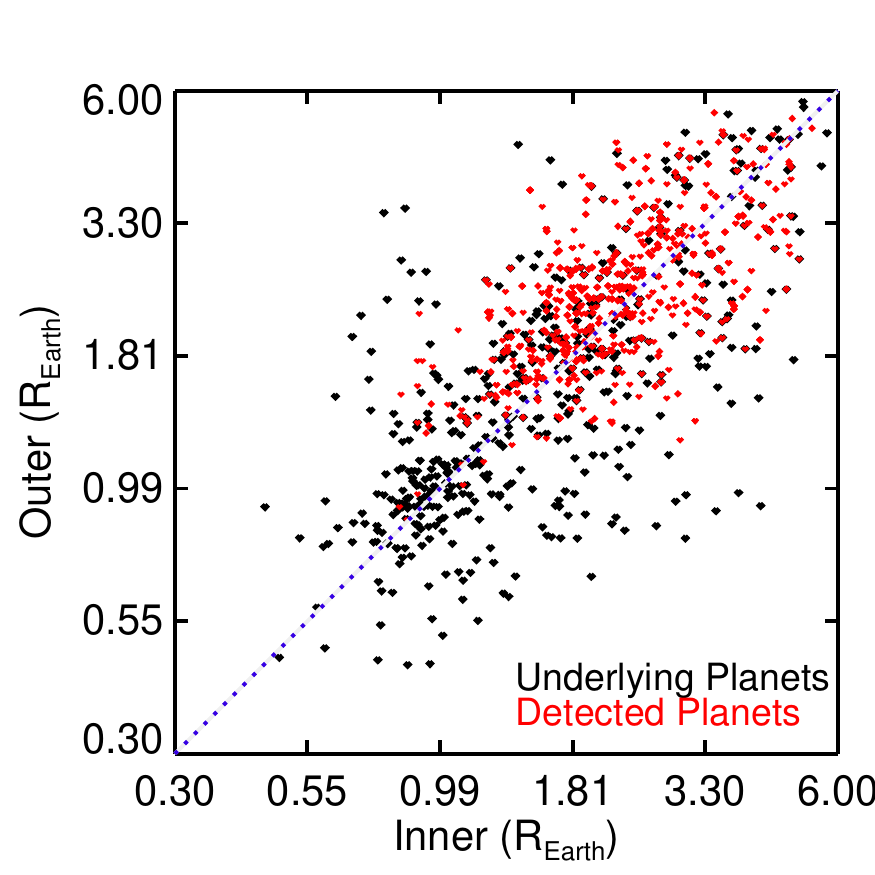}
    \includegraphics[width=0.47\textwidth]{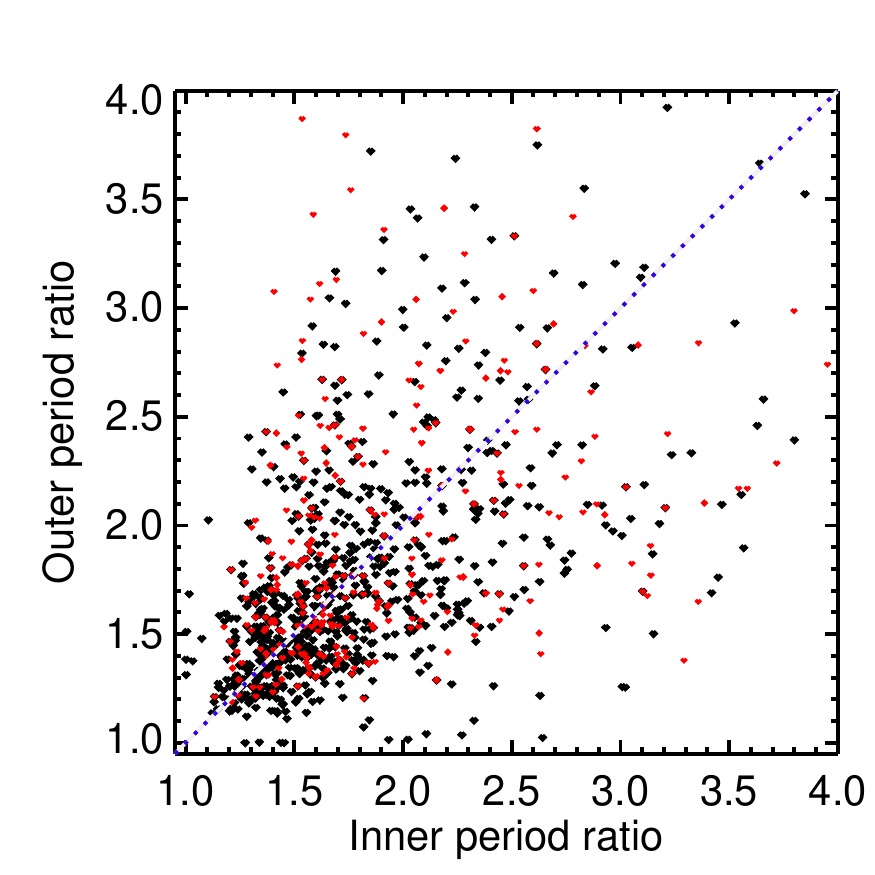}
     \caption{Adjacent planets in simulated systems exhibit similarity in their size (top) and spacing (bottom). Black points indicated underlying simulated planets and red points are forward modeled simulated detected transiting planets.}
    \label{fig:peas_corr}
\end{figure}

\citet{mill17} and \citet{weis18} found that \kep planets within the same system exhibit similar sizes and spacings, and there has been controversy over whether this ``peas in a pod'' pattern is astrophysical or a selection effect \citep{zhu19,weis19}. In Fig. \ref{fig:peas_corr}, we plot the size of outer vs. inner planets in transiting pairs and outer vs. inner period ratio in transiting triplets. We find that our simulated planets exhibit intra-system similarity and that the similarity is only slightly enhanced by selection effects. The correlation between outer vs. inner period ratio less tight than the correlation between adjacent radii and is most evident by eye in the lack of planets in the top left and bottom right corners of the plot (Fig. \ref{fig:peas_corr}, bottom panel). For outer vs. inner period ratio, the Pearson R coefficient is 0.46 ($p=4 \times 10^{-15}$) for the 450 plotted underlying simulated planets and 0.34 ($p=1.3 \times 10^{-13}$) for the 450 plotted forward modeled simulated detected transiting planets.

\begin{figure*}
    \centering
        \includegraphics[width=0.47\textwidth]{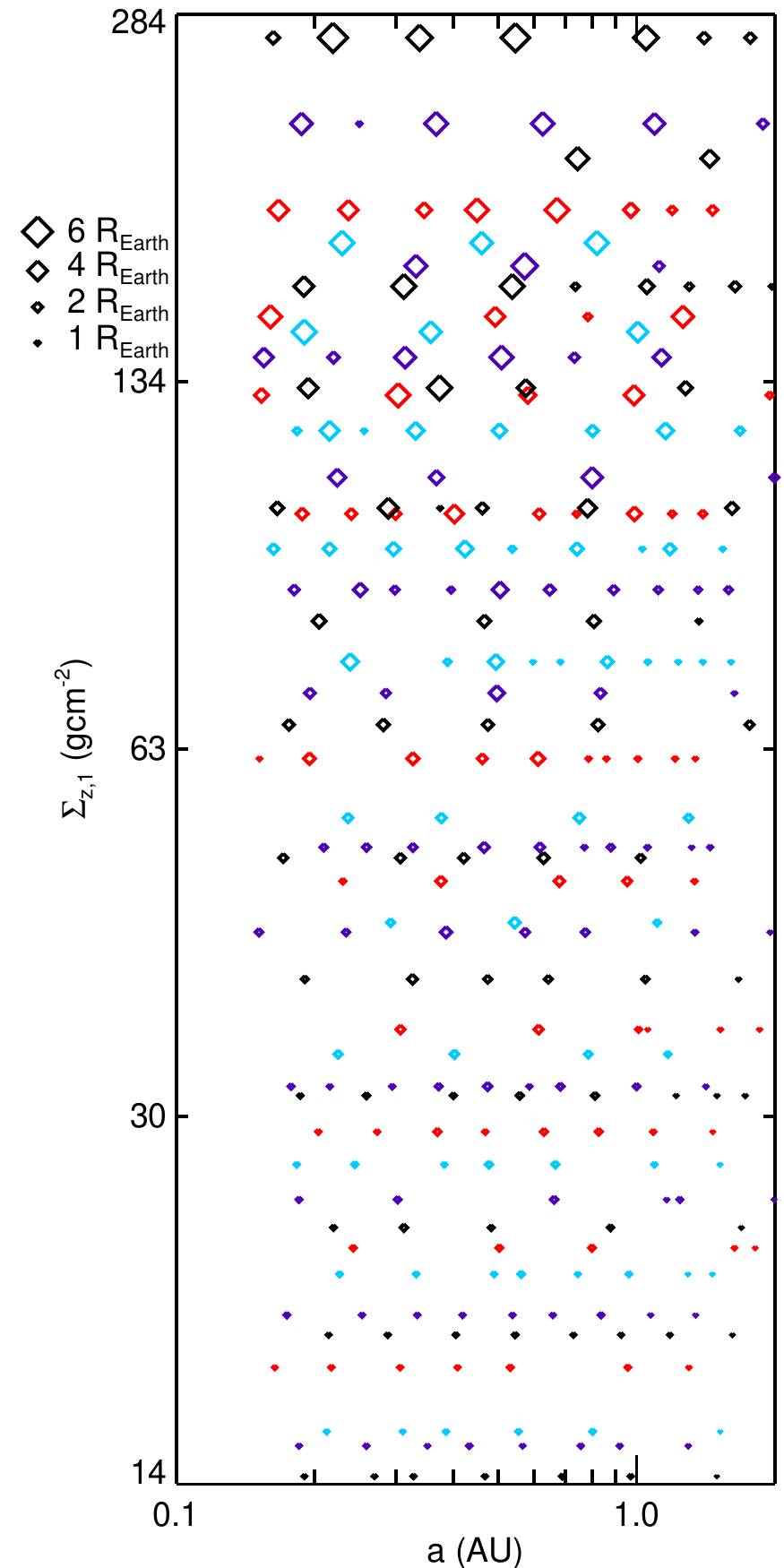}
    \includegraphics[width=0.47\textwidth]{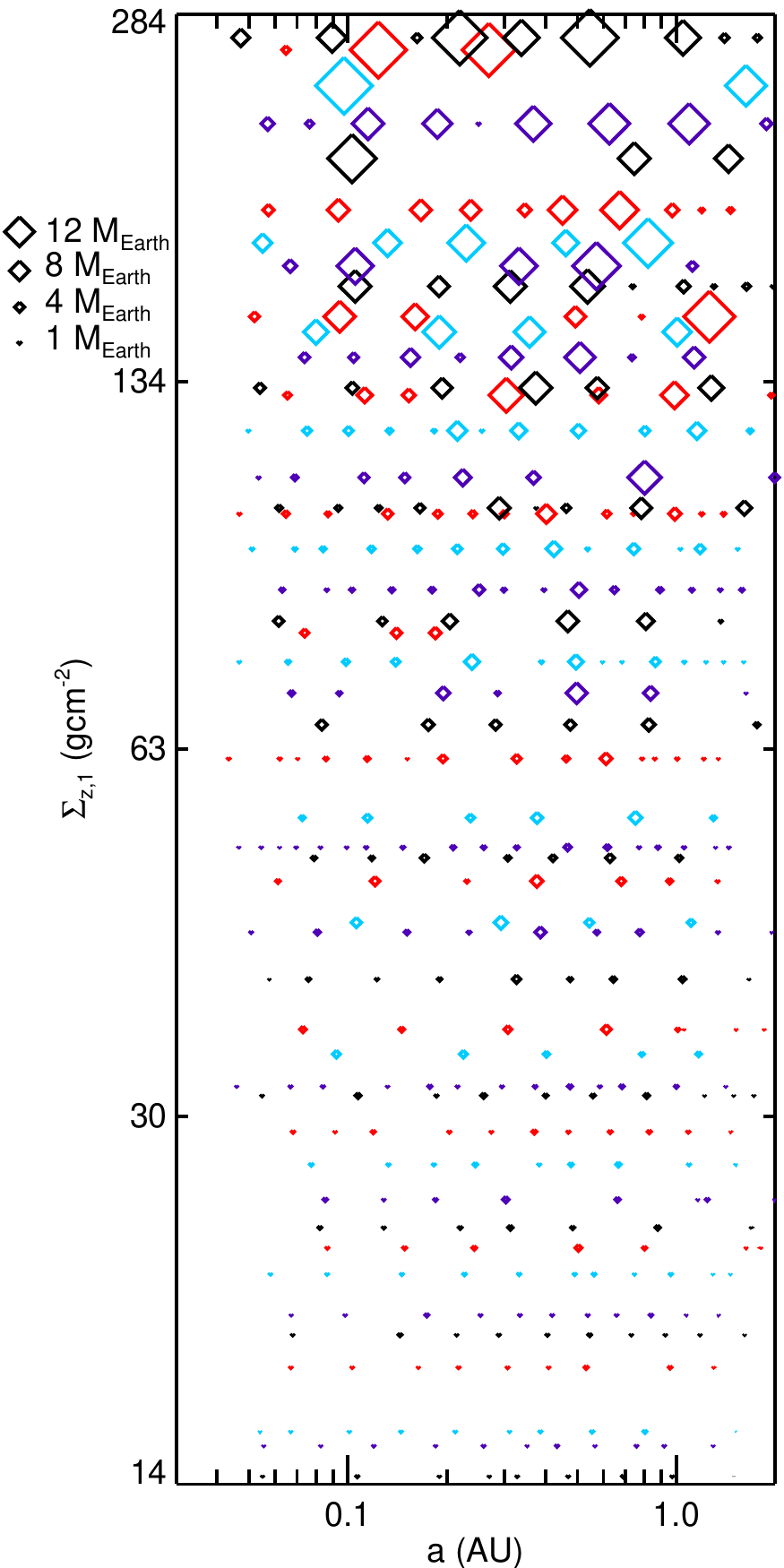}
     \caption{Planets within each simulated system exhibit similarity in their size (left) and mass (right). We plot 50 randomly selected simulations, sorted by the amount of solids in the disk (parametrized as $\sigz$, Section \ref{subsec:disk}).}
    \label{fig:peas}
\end{figure*}

Despite the stochasticity in growth of masses and radii introduced by giant impacts and subsequent atmospheric loss via giant impacts, our simulated planets have similarities in their sizes and spacings that are ultimately controlled by the amount of solids in the formation region (our $\sigz$). Fig. \ref{fig:peas} shows planets in 50 randomly selected simulated underlying systems: systems range from smaller masses and radii with somewhat larger spacings at small $\sigz$ to larger masses and radii with more compact spacings at large $\sigz$. A clustering in mass-radius space -- as seen by \citet{mill17} in real systems - is evident in the bottom panel of our Fig. \ref{fig:mr}: each range $\sigz$ occupies different parts of mass-radius space. 

We find our simulated systems are generally self-similar in mass and period across orbital period beyond 0.15 AU. Within 0.15 AU, masses tend to be smaller and, as noted by DLC16, spacings tend to be larger (which they attribute to the more massive outer embryos stirring up those at short orbital periods. (As noted earlier, we do not consider planetary radii within 0.15 AU because we neglect photoevaporation.)

In contrast, the alternative, steeper disk profiles we consider in Section \ref{subsec:other} produce significant trends of larger masses and radii at shorter orbital periods, even beyond 0.15 AU. Within a given system, this alternative disk profile produces larger, more massive planets on the inside and smaller, lower mass planets on the outside. However, the diversity within a system is small compared to the diversity among systems. Planets formed from steeper disk profiles do exhibit intra-system similarity according to the correlation metrics of \citet{weis18} and distance metric of \citet{mill17} but may be at odds with the observed increase in mass/size with orbital period.

Although one might expect post-gas atmospheric loss via collisions to destroy intra-system uniformity, we find that our results are not sensitive to the atmospheric loss treatment (Section \ref{subsec:sims}). Systems tend to either finish growth during the gas disk stage and not experience much atmospheric loss or to grow significantly after the gas disks age and have most planets experience significant loss. The most significant effect is that our most extreme treatment of atmospheric loss (in which the planet loses its entire atmosphere in a single collision) amplifies a trend of producing smaller planets beyond 1 AU, where dynamical timescales are long and cores need more time to form. Since there are very few transiting planets beyond 1 AU, this effect is only noticeable when examining the underlying systems.

\section{Conclusion} \label{sec:conclusion}

We investigated whether a continuum of formation conditions can account for the diversity of orbital and compositional properties observed for super Earths, including the apparent dichotomy between single transiting and multiple transiting system. We focused on intrinsic variation in the amount of solids delivered from the outer disk to form super-Earths in situ close to the star. We found that this variation in disk conditions can account for super-Earths ranging from rocky to gas-enveloped and from ``dynamically cold'' (tight, flat, compact, and circular) to ``dynamically hot'' (wider spaced, more elliptical and inclined). The continuum of formation conditions can account for the diversity in \kep planets' orbital period ratios, transit duration ratios, and transit multiplicity (Section \ref{subsec:orbresults}); higher eccentricities for single than multi transiting planets (Section \ref{subsec:ei}; but note that our models cannot explain the most elliptical singles with $e>0.3$); smaller eccentricities for larger planets (and vice versa) among pairs that exhibit transit timing variations (Section \ref{subsec:size}); scatter in the mass-radius relation, including lower densities for planets with masses measured with TTVs than RVs (Section \ref{subsec:mr}); and similarity in planets' sizes and spacings within each system (Section \ref{subsec:peas}). 

In our simulations, planets form via giant impacts in a depleted gas disk and undergo subsequent gas-free orbital evolution. When little solid material is available for in situ formation -- like in our Solar System -- planets' growth is stalled during the gas disk stage, and they finish forming via giant impacts after the gas disk dissipates. Such planets end up rocky and more dynamically hot, like our Solar System's terrestrial planets. When more material is available, planets can finish forming via giant impacts during the gas stage, accrete envelopes, and remain in circular, flat, and compact orbital configurations. Our simulated underlying population is weighted toward disks with less material available like the Solar System. However, disks with more solids account for a greater share of the simulated detected transiting planets and an even greater share of multi-transiting systems. At an extreme, too much solid material also leads to dynamically hot systems, because damping from the depleted gas disk is insufficient to avoid planets exciting eccentricities and mutual inclinations amongst themselves. However, in practice these conditions create planets that are much more massive than those observed.

In addition to the amount of solids, we explored how the level of gas depletion at the end of the disk lifetime and radial distribution of solids affect the \kep planets' period ratios, Hill spacing, duration ratios, and transit multiplicity. Consistent with DLC16, we found that too much or little disk gas does not allow planets to finish their giant impacts during the gas disk stage, leading to systems that are too dynamically hot and, consistent with MB16, that steeper radial profiles lead to high transit multiplicity. In the future, we plan a more complete exploration of disk parameters using Gaussian process emulation to keep the number of simulations computationally feasible. We can also explore more sophisticated disk conditions like non-smooth radial profiles (e.g., gaps, pile ups), continuous replenishment of solids from the outer disks, embryo masses established by pebble accretion, and a range of collisional outcomes besides accretion. Furthermore, we can investigate a range disk dissipation timescales and functional forms. Disk dissipation timescales may vary from disk to disk, which could also contribute\footnote{However, we expect intrinsic variation in the amount of solids to dominate. In low solid surface density disks, the planets' growth is stalled due to damping from gas, so we expect longer disk dissipation timescales would still not allow them to finish forming the presence of gas. However, we expect variations in dissipation timescale could contribute significantly to the scatter in the mass-radius relationship by allowing certain planets more or less time to accrete.} to the diversity in planetary properties.

There are several other avenues for future work. TESS systems will provide another test ground for theories of planet formation, particularly because radial-velocity follow up will enable distinguishing the underlying architectures in systems where not all planets transit (e.g., \citealt{ball19}). Including photoevaporation in our models would allow to compare our radii to those observed at shorter orbital periods, and to assess whether mass-radius trends are primarily driven by formation or evolution. Including giant planets in a subset of simulations would allow us to assess their relative contribution to the diversity of planetary properties. Compared to the observed distribution with an assumed mass-radius relationship, we found a shortage of planets with Hill spacings of $\Delta < 15$ (Section \ref{subsec:orbresults}). A more complete exploration of disk parameters combined with more mass measurements of observed planets will clarify whether this discrepancy results from assumed mass-radius relationship (in particular, the assumption that it is constant with orbital period ratio, which is not born out in our simulations: Section \ref{subsec:mr}), the need to tune the disk gas depletion factor to tighter than an order of magnitude, or a fundamental inability of in situ formation to form massive planets with tight period ratios.

Planet formation models have many knobs to turn, raising the concern that they can ``post-dict'' any observable result yet not give us true insight into which physical processes establish the surprising diversity among inner solar systems. We have chosen to focus on solid surface density as a property we know must vary among disks, if in situ formation is responsible for systems ranging from the Solar System to Kepler-11. Our findings -- that a continuum in this single property can account for the observed orbital and compositional diversity -- support the in situ formation paradigm and the idea that planet properties are significantly sculpted by nature (formation) in addition to nurture (evolution).

\begin{acknowledgments}
We thank Eric Ford, Matthias He, Danley Hsu, Lauren Weiss, Angie Wolfgang, and Jason Wright for helpful discussions and comments. We thank the referees for helpful comments that improved the paper. MGM acknowledges that this material is based upon work supported by the National Science Foundation Graduate Research Fellowship Program under Grant No. DGE1255832. Any opinions, findings, and conclusions or recommendations expressed in this material are those of the author and do not necessarily reflect the views of the National Science Foundation. RID and SJM gratefully acknowledge support from NASA XRP 80NSSC18K0355. SJM also acknowledges support from the Missouri State University Graduate College. AK acknowledges support from the 2015 Summer Scholars Fund of the Marian E. Koshland Integrated Natural Sciences Center at Haverford College. We gratefully acknowledge support from the Alfred P. Sloan Foundation's Sloan Research Fellowship. The Center for Exoplanets and Habitable Worlds is supported by the Pennsylvania State University, the Eberly College of Science, and the Pennsylvania Space Grant Consortium. This research was supported in part by the National Science Foundation under Grant No. NSF PHY-1748958. The valuable collection of planet candidates were discovered by NASA’s Kepler mission and compiled from NASA’s Exoplanet Archive, which is operated by the California Institute of Technology, under contract with the National Aeronautics and Space Administration under the Exoplanet Exploration Program. Funding for the Kepler Discovery Mission is provided by NASA's Science Mission Directorate.
\end{acknowledgments}

\appendix

We empirically explore the detection efficiency of the \kep pipeline by comparing candidate lists \citep{Batalha2013,Burke2014,Mullally2015,rowe15,coug16} that were released throughout the \kep Mission. We assess the detection efficiency of candidates as a function of their multiple event statistic (MES), which is an estimate of total signal-to-noise ratio over all transits \citep{jenk02}. We identify ``missing'' candidates as those present in a later candidate list but absent in an earlier one despite our inference (based on number of transits) that they had $MES>7.1$ in the earlier list. For example, imagine a candidate is detected on the quarter 1--12 (Q1--12) list with $MES=20$ and, based on the number of transits in Q1--Q8, we calculate that it should have had $MES=17$ on the Q1--Q8 list. If the candidate is not included on the Q1--Q8 list, we consider it missing. 

\begin{figure}
    \centering
        \includegraphics[width=0.70\textwidth]{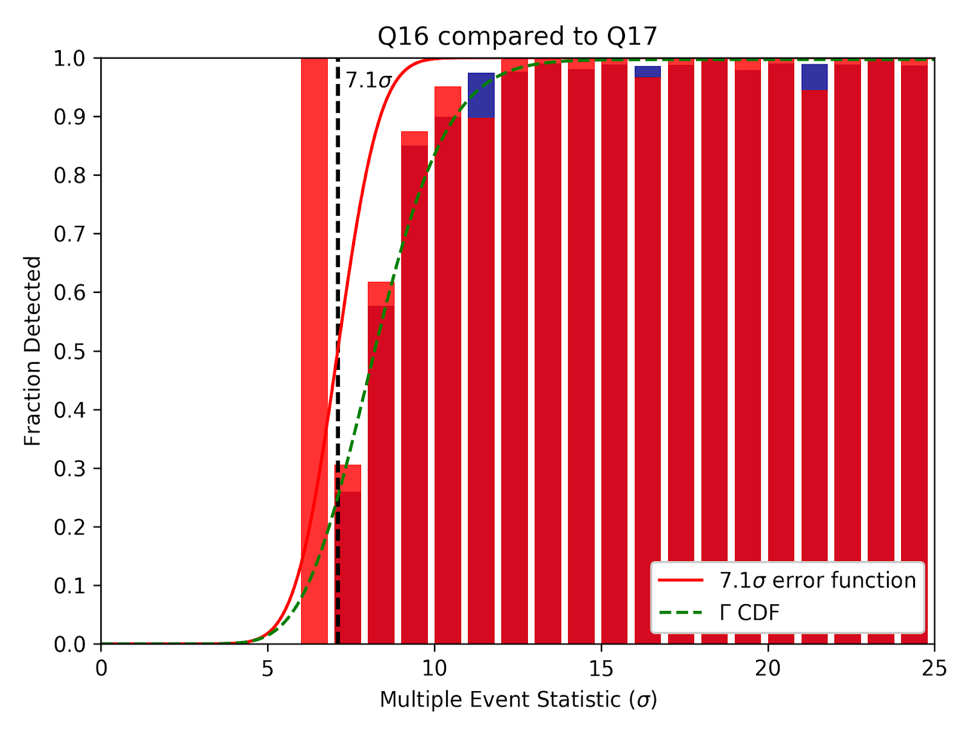}
     \caption{The detection efficiency of the Kepler candidate lists. The blue histogram shows the detection efficiency for candidates with periods less than 40 days, while the red histogram shows the detection efficiency for periods greater than 40 days. The black dashed line is the detection threshold of $MES>7.1$.The solid red line is the hypothetical performance of the pipeline given perfectly whitened noise, which is an error function centered on $MES=7.1$. The green dashed line is the cumulative gamma distribution function fitted in \citet{chri16} from injection and recovery tests. This figure contains detection efficiencies for $MES < 7.1$; some (detected) candidates in the earlier list have a calculated $MES < 7.1$, while by design no missing candidates have $MES < 7.1$.}
    \label{fig:mes}
\end{figure}

We investigate whether ``missing'' candidates exhibit trends in orbital period or impact parameter that might compromise our comparison to the observed sample with our forward modeling process. In earlier lists, we find that candidates are more likely to be missing at a given $MES$ if they have large impact parameters or long orbital periods. However, by Q1--Q16 \citep{rowe15} vs. Q1--Q17 \citep{coug16}, this trend has disappeared. In Fig. \ref{fig:mes}, we plot the fraction of candidates detected in Q1--Q16 as a function of MES inferred from Q1--Q17. The fraction is consistent with \citet{chri16}'s empirical detection efficiency function computed from injection and recovery tests and shows no trend with orbital period.

\newpage

\bibliographystyle{apj}
\bibliography{main}

\begin{thebibliography}{}
\expandafter\ifx\csname natexlab\endcsname\relax\def\natexlab#1{#1}\fi

\bibitem[{{Ballard}(2019)}]{ball19}
{Ballard}, S. 2019, \aj, 157, 113

\bibitem[{{Ballard} \& {Johnson}(2016)}]{ball16}
{Ballard}, S., \& {Johnson}, J.~A. 2016, \apj, 816, 66

\bibitem[{{Batalha} {et~al.}(2013){Batalha}, {Rowe}, {Bryson}, {Barclay},
  {Burke}, {Caldwell}, {Christiansen}, {Mullally}, {Thompson}, {Brown},
  {Dupree}, {Fabrycky}, {Ford}, {Fortney}, {Gilliland}, {Isaacson}, {Latham},
  {Marcy}, {Quinn}, {Ragozzine}, {Shporer}, {Borucki}, {Ciardi}, {Gautier},
  {Haas}, {Jenkins}, {Koch}, {Lissauer}, {Rapin}, {Basri}, {Boss}, {Buchhave},
  {Carter}, {Charbonneau}, {Christensen-Dalsgaard}, {Clarke}, {Cochran},
  {Demory}, {Desert}, {Devore}, {Doyle}, {Esquerdo}, {Everett}, {Fressin},
  {Geary}, {Girouard}, {Gould}, {Hall}, {Holman}, {Howard}, {Howell},
  {Ibrahim}, {Kinemuchi}, {Kjeldsen}, {Klaus}, {Li}, {Lucas}, {Meibom},
  {Morris}, {Pr{\v s}a}, {Quintana}, {Sanderfer}, {Sasselov}, {Seader},
  {Smith}, {Steffen}, {Still}, {Stumpe}, {Tarter}, {Tenenbaum}, {Torres},
  {Twicken}, {Uddin}, {Van Cleve}, {Walkowicz}, \& {Welsh}}]{Batalha2013}
{Batalha}, N.~M., {Rowe}, J.~F., {Bryson}, S.~T., {et~al.} 2013, \apjs, 204, 24

\bibitem[{{Borucki} {et~al.}(2011){Borucki}, {Koch}, {Basri}, {Batalha},
  {Boss}, {Brown}, {Caldwell}, {Christensen-Dalsgaard}, {Cochran}, {DeVore},
  {Dunham}, {Dupree}, {Gautier}, {Geary}, {Gilliland}, {Gould}, {Howell},
  {Jenkins}, {Kjeldsen}, {Latham}, {Lissauer}, {Marcy}, {Monet}, {Sasselov},
  {Tarter}, {Charbonneau}, {Doyle}, {Ford}, {Fortney}, {Holman}, {Seager},
  {Steffen}, {Welsh}, {Allen}, {Bryson}, {Buchhave}, {Chandrasekaran},
  {Christiansen}, {Ciardi}, {Clarke}, {Dotson}, {Endl}, {Fischer}, {Fressin},
  {Haas}, {Horch}, {Howard}, {Isaacson}, {Kolodziejczak}, {Li}, {MacQueen},
  {Meibom}, {Prsa}, {Quintana}, {Rowe}, {Sherry}, {Tenenbaum}, {Torres},
  {Twicken}, {Van Cleve}, {Walkowicz}, \& {Wu}}]{Borucki2011}
{Borucki}, W.~J., {Koch}, D.~G., {Basri}, G., {et~al.} 2011, \apj, 728, 117

\bibitem[{{Brakensiek} \& {Ragozzine}(2016)}]{Brakensiek2016}
{Brakensiek}, J., \& {Ragozzine}, D. 2016, \apj, 821, 47

\bibitem[{{Burke} {et~al.}(2014){Burke}, {Bryson}, {Mullally}, {Rowe},
  {Christiansen}, {Thompson}, {Coughlin}, {Haas}, {Batalha}, {Caldwell},
  {Jenkins}, {Still}, {Barclay}, {Borucki}, {Chaplin}, {Ciardi}, {Clarke},
  {Cochran}, {Demory}, {Esquerdo}, {Gautier}, {Gilliland}, {Girouard}, {Havel},
  {Henze}, {Howell}, {Huber}, {Latham}, {Li}, {Morehead}, {Morton}, {Pepper},
  {Quintana}, {Ragozzine}, {Seader}, {Shah}, {Shporer}, {Tenenbaum}, {Twicken},
  \& {Wolfgang}}]{Burke2014}
{Burke}, C.~J., {Bryson}, S.~T., {Mullally}, F., {et~al.} 2014, \apjs, 210, 19

\bibitem[{{Burke} {et~al.}(2015){Burke}, {Christiansen}, {Mullally}, {Seader},
  {Huber}, {Rowe}, {Coughlin}, {Thompson}, {Catanzarite}, {Clarke}, {Morton},
  {Caldwell}, {Bryson}, {Haas}, {Batalha}, {Jenkins}, {Tenenbaum}, {Twicken},
  {Li}, {Quintana}, {Barclay}, {Henze}, {Borucki}, {Howell}, \&
  {Still}}]{Burke2015}
{Burke}, C.~J., {Christiansen}, J.~L., {Mullally}, F., {et~al.} 2015, \apj,
  809, 8

\bibitem[{{Chambers}(2001)}]{cham01}
{Chambers}, J.~E. 2001, \icarus, 152, 205

\bibitem[{{Chambers} {et~al.}(1996){Chambers}, {Wetherill}, \&
  {Boss}}]{Chambers1996}
{Chambers}, J.~E., {Wetherill}, G.~W., \& {Boss}, A.~P. 1996, \icarus, 119, 261

\bibitem[{{Chiang} \& {Laughlin}(2013)}]{Chiang2013}
{Chiang}, E., \& {Laughlin}, G. 2013, \mnras, 431, 3444

\bibitem[{{Chiang} \& {Youdin}(2010)}]{chian10}
{Chiang}, E., \& {Youdin}, A.~N. 2010, Annual Review of Earth and Planetary
  Sciences, 38, 493

\bibitem[{{Christiansen} {et~al.}(2015){Christiansen}, {Clarke}, {Burke},
  {Seader}, {Jenkins}, {Twicken}, {Catanzarite}, {Smith}, {Batalha}, {Haas},
  {Thompson}, {Campbell}, {Sabale}, \& {Kamal Uddin}}]{Christiansen2015}
{Christiansen}, J.~L., {Clarke}, B.~D., {Burke}, C.~J., {et~al.} 2015, \apj,
  810, 95

\bibitem[{{Christiansen} {et~al.}(2016){Christiansen}, {Clarke}, {Burke},
  {Jenkins}, {Bryson}, {Coughlin}, {Mullally}, {Thompson}, {Twicken},
  {Batalha}, {Haas}, {Catanzarite}, {Campbell}, {Kamal Uddin}, {Zamudio},
  {Smith}, \& {Henze}}]{chri16}
---. 2016, \apj, 828, 99

\bibitem[{{Coughlin} {et~al.}(2016){Coughlin}, {Mullally}, {Thompson}, {Rowe},
  {Burke}, {Latham}, {Batalha}, {Ofir}, {Quarles}, {Henze}, {Wolfgang},
  {Caldwell}, {Bryson}, {Shporer}, {Catanzarite}, {Akeson}, {Barclay},
  {Borucki}, {Boyajian}, {Campbell}, {Christiansen}, {Girouard}, {Haas},
  {Howell}, {Huber}, {Jenkins}, {Li}, {Patil-Sabale}, {Quintana}, {Ramirez},
  {Seader}, {Smith}, {Tenenbaum}, {Twicken}, \& {Zamudio}}]{coug16}
{Coughlin}, J.~L., {Mullally}, F., {Thompson}, S.~E., {et~al.} 2016, \apjs,
  224, 12

\bibitem[{{Dawson} {et~al.}(2015){Dawson}, {Chiang}, \& {Lee}}]{Dawson2015}
{Dawson}, R.~I., {Chiang}, E., \& {Lee}, E.~J. 2015, \mnras, 453, 1471

\bibitem[{{Dawson} {et~al.}(2016){Dawson}, {Lee}, \& {Chiang}}]{Dawson2016}
{Dawson}, R.~I., {Lee}, E.~J., \& {Chiang}, E. 2016, \apj, 822, 54

\bibitem[{{Deck} \& {Batygin}(2015)}]{Deck2015}
{Deck}, K.~M., \& {Batygin}, K. 2015, \apj, 810, 119

\bibitem[{{Dressing} {et~al.}(2015){Dressing}, {Charbonneau}, {Dumusque},
  {Gettel}, {Pepe}, {Collier Cameron}, {Latham}, {Molinari}, {Udry}, {Affer},
  {Bonomo}, {Buchhave}, {Cosentino}, {Figueira}, {Fiorenzano}, {Harutyunyan},
  {Haywood}, {Johnson}, {Lopez-Morales}, {Lovis}, {Malavolta}, {Mayor},
  {Micela}, {Motalebi}, {Nascimbeni}, {Phillips}, {Piotto}, {Pollacco},
  {Queloz}, {Rice}, {Sasselov}, {S{\'e}gransan}, {Sozzetti}, {Szentgyorgyi}, \&
  {Watson}}]{Dressing2015}
{Dressing}, C.~D., {Charbonneau}, D., {Dumusque}, X., {et~al.} 2015, \apj, 800,
  135

\bibitem[{{Fabrycky} {et~al.}(2014){Fabrycky}, {Lissauer}, {Ragozzine}, {Rowe},
  {Steffen}, {Agol}, {Barclay}, {Batalha}, {Borucki}, {Ciardi}, {Ford},
  {Gautier}, {Geary}, {Holman}, {Jenkins}, {Li}, {Morehead}, {Morris},
  {Shporer}, {Smith}, {Still}, \& {Van Cleve}}]{Fabrycky2014}
{Fabrycky}, D.~C., {Lissauer}, J.~J., {Ragozzine}, D., {et~al.} 2014, \apj,
  790, 146

\bibitem[{{Fang} \& {Margot}(2012)}]{Fang2012}
{Fang}, J., \& {Margot}, J.-L. 2012, \apj, 761, 92

\bibitem[{{Fang} \& {Margot}(2013)}]{fang13}
---. 2013, \apj, 767, 115

\bibitem[{{Ford} \& {Chiang}(2007)}]{Ford2007}
{Ford}, E.~B., \& {Chiang}, E.~I. 2007, \apj, 661, 602

\bibitem[{{Fressin} {et~al.}(2013){Fressin}, {Torres}, {Charbonneau}, {Bryson},
  {Christiansen}, {Dressing}, {Jenkins}, {Walkowicz}, \&
  {Batalha}}]{Fressin2013}
{Fressin}, F., {Torres}, G., {Charbonneau}, D., {et~al.} 2013, \apj, 766, 81

\bibitem[{{Hadden} \& {Lithwick}(2014)}]{Hadden2014}
{Hadden}, S., \& {Lithwick}, Y. 2014, \apj, 787, 80

\bibitem[{{Hansen} \& {Murray}(2013)}]{Hansen2013}
{Hansen}, B.~M.~S., \& {Murray}, N. 2013, \apj, 775, 53

\bibitem[{{He} {et~al.}(2019){He}, {Ford}, \& {Ragozzine}}]{he19}
{He}, M.~Y., {Ford}, E.~B., \& {Ragozzine}, D. 2019, \mnras, 490, 4575

\bibitem[{{Howard} {et~al.}(2010){Howard}, {Marcy}, {Johnson}, {Fischer},
  {Wright}, {Isaacson}, {Valenti}, {Anderson}, {Lin}, \& {Ida}}]{Howard2010}
{Howard}, A.~W., {Marcy}, G.~W., {Johnson}, J.~A., {et~al.} 2010, Science, 330,
  653

\bibitem[{{Huang} {et~al.}(2017){Huang}, {Petrovich}, \& {Deibert}}]{Huang2017}
{Huang}, C.~X., {Petrovich}, C., \& {Deibert}, E. 2017, \aj, 153, 210

\bibitem[{{Inamdar} \& {Schlichting}(2015)}]{inam15}
{Inamdar}, N.~K., \& {Schlichting}, H.~E. 2015, \mnras, 448, 1751

\bibitem[{{Izidoro} {et~al.}(2017){Izidoro}, {Ogihara}, {Raymond},
  {Morbidelli}, {Pierens}, {Bitsch}, {Cossou}, \& {Hersant}}]{izid17}
{Izidoro}, A., {Ogihara}, M., {Raymond}, S.~N., {et~al.} 2017, \mnras, 470,
  1750

\bibitem[{{Jenkins}(2002)}]{jenk02}
{Jenkins}, J.~M. 2002, \apj, 575, 493

\bibitem[{{Johansen} {et~al.}(2012){Johansen}, {Davies}, {Church}, \&
  {Holmelin}}]{joha12}
{Johansen}, A., {Davies}, M.~B., {Church}, R.~P., \& {Holmelin}, V. 2012, \apj,
  758, 39

\bibitem[{{Kokubo} \& {Ida}(1998)}]{Kokubo1998}
{Kokubo}, E., \& {Ida}, S. 1998, \icarus, 131, 171

\bibitem[{{Kominami} \& {Ida}(2002)}]{Kominami2002}
{Kominami}, J., \& {Ida}, S. 2002, \icarus, 157, 43

\bibitem[{{Lee}(2019)}]{Lee19}
{Lee}, E.~J. 2019, \apj, 878, 36

\bibitem[{{Lee} \& {Chiang}(2016)}]{Lee2016}
{Lee}, E.~J., \& {Chiang}, E. 2016, \apj, 817, 90

\bibitem[{{Lee} {et~al.}(2014){Lee}, {Chiang}, \& {Ormel}}]{lee14}
{Lee}, E.~J., {Chiang}, E., \& {Ormel}, C.~W. 2014, \apj, 797, 95

\bibitem[{{Lissauer} {et~al.}(2011){Lissauer}, {Ragozzine}, {Fabrycky},
  {Steffen}, {Ford}, {Jenkins}, {Shporer}, {Holman}, {Rowe}, {Quintana},
  {Batalha}, {Borucki}, {Bryson}, {Caldwell}, {Carter}, {Ciardi}, {Dunham},
  {Fortney}, {Gautier}, {Howell}, {Koch}, {Latham}, {Marcy}, {Morehead}, \&
  {Sasselov}}]{Lissauer2011}
{Lissauer}, J.~J., {Ragozzine}, D., {Fabrycky}, D.~C., {et~al.} 2011, \apjs,
  197, 8

\bibitem[{{Lopez} \& {Fortney}(2014)}]{Lopez14}
{Lopez}, E.~D., \& {Fortney}, J.~J. 2014, \apj, 792, 1

\bibitem[{{Malhotra}(2015)}]{Malhotra2015}
{Malhotra}, R. 2015, \apj, 808, 71

\bibitem[{{Millholland} {et~al.}(2017){Millholland}, {Wang}, \&
  {Laughlin}}]{mill17}
{Millholland}, S., {Wang}, S., \& {Laughlin}, G. 2017, \apjl, 849, L33

\bibitem[{{Mills} {et~al.}(2019){Mills}, {Howard}, {Petigura}, {Fulton},
  {Isaacson}, \& {Weiss}}]{mills19}
{Mills}, S.~M., {Howard}, A.~W., {Petigura}, E.~A., {et~al.} 2019, \aj, 157,
  198

\bibitem[{{Moriarty} \& {Ballard}(2016)}]{Moriarty2016}
{Moriarty}, J., \& {Ballard}, S. 2016, \apj, 832, 34

\bibitem[{{Morton} \& {Johnson}(2011)}]{Morton2011}
{Morton}, T.~D., \& {Johnson}, J.~A. 2011, \apj, 738, 170

\bibitem[{{Mullally} {et~al.}(2015){Mullally}, {Coughlin}, {Thompson}, {Rowe},
  {Burke}, {Latham}, {Batalha}, {Bryson}, {Christiansen}, {Henze}, {Ofir},
  {Quarles}, {Shporer}, {Van Eylen}, {Van Laerhoven}, {Shah}, {Wolfgang},
  {Chaplin}, {Xie}, {Akeson}, {Argabright}, {Bachtell}, {Barclay}, {Borucki},
  {Caldwell}, {Campbell}, {Catanzarite}, {Cochran}, {Duren}, {Fleming},
  {Fraquelli}, {Girouard}, {Haas}, {He{\l}miniak}, {Howell}, {Huber}, {Larson},
  {Gautier}, {Jenkins}, {Li}, {Lissauer}, {McArthur}, {Miller}, {Morris},
  {Patil-Sabale}, {Plavchan}, {Putnam}, {Quintana}, {Ramirez}, {Silva Aguirre},
  {Seader}, {Smith}, {Steffen}, {Stewart}, {Stober}, {Still}, {Tenenbaum},
  {Troeltzsch}, {Twicken}, \& {Zamudio}}]{Mullally2015}
{Mullally}, F., {Coughlin}, J.~L., {Thompson}, S.~E., {et~al.} 2015, \apjs,
  217, 31

\bibitem[{{Owen} {et~al.}(2012){Owen}, {Clarke}, \& {Ercolano}}]{owen12}
{Owen}, J.~E., {Clarke}, C.~J., \& {Ercolano}, B. 2012, \mnras, 422, 1880

\bibitem[{{Owen} {et~al.}(2011){Owen}, {Ercolano}, \& {Clarke}}]{owen11}
{Owen}, J.~E., {Ercolano}, B., \& {Clarke}, C.~J. 2011, \mnras, 412, 13

\bibitem[{{Papaloizou} \& {Larwood}(2000)}]{Papaloizou2000}
{Papaloizou}, J.~C.~B., \& {Larwood}, J.~D. 2000, \mnras, 315, 823

\bibitem[{{Petrovich} {et~al.}(2014){Petrovich}, {Tremaine}, \&
  {Rafikov}}]{Petrovich2014}
{Petrovich}, C., {Tremaine}, S., \& {Rafikov}, R. 2014, \apj, 786, 101

\bibitem[{{Pu} \& {Wu}(2015)}]{Pu2015}
{Pu}, B., \& {Wu}, Y. 2015, \apj, 807, 44

\bibitem[{{Rein}(2012)}]{Rein2012}
{Rein}, H. 2012, \mnras, 427, L21

\bibitem[{{Rowe} {et~al.}(2015){Rowe}, {Coughlin}, {Antoci}, {Barclay},
  {Batalha}, {Borucki}, {Burke}, {Bryson}, {Caldwell}, {Campbell},
  {Catanzarite}, {Christiansen}, {Cochran}, {Gilliland}, {Girouard}, {Haas},
  {He{\l}miniak}, {Henze}, {Hoffman}, {Howell}, {Huber}, {Hunter},
  {Jang-Condell}, {Jenkins}, {Klaus}, {Latham}, {Li}, {Lissauer}, {McCauliff},
  {Morris}, {Mullally}, {Ofir}, {Quarles}, {Quintana}, {Sabale}, {Seader},
  {Shporer}, {Smith}, {Steffen}, {Still}, {Tenenbaum}, {Thompson}, {Twicken},
  {Van Laerhoven}, {Wolfgang}, \& {Zamudio}}]{rowe15}
{Rowe}, J.~F., {Coughlin}, J.~L., {Antoci}, V., {et~al.} 2015, \apjs, 217, 16

\bibitem[{{Schlichting}(2014)}]{Schlichting14}
{Schlichting}, H.~E. 2014, \apjl, 795, L15

\bibitem[{{Spalding} \& {Batygin}(2016)}]{Spalding2016}
{Spalding}, C., \& {Batygin}, K. 2016, \apj, 830, 5

\bibitem[{{Thompson} {et~al.}(2018){Thompson}, {Coughlin}, {Hoffman},
  {Mullally}, {Christiansen}, {Burke}, {Bryson}, {Batalha}, {Haas},
  {Catanzarite}, {Rowe}, {Barentsen}, {Caldwell}, {Clarke}, {Jenkins}, {Li},
  {Latham}, {Lissauer}, {Mathur}, {Morris}, {Seader}, {Smith}, {Klaus},
  {Twicken}, {Van Cleve}, {Wohler}, {Akeson}, {Ciardi}, {Cochran}, {Henze},
  {Howell}, {Huber}, {Pr{\v{s}}a}, {Ram{\'\i}rez}, {Morton}, {Barclay},
  {Campbell}, {Chaplin}, {Charbonneau}, {Christensen-Dalsgaard}, {Dotson},
  {Doyle}, {Dunham}, {Dupree}, {Ford}, {Geary}, {Girouard}, {Isaacson},
  {Kjeldsen}, {Quintana}, {Ragozzine}, {Shabram}, {Shporer}, {Silva Aguirre},
  {Steffen}, {Still}, {Tenenbaum}, {Welsh}, {Wolfgang}, {Zamudio}, {Koch}, \&
  {Borucki}}]{thom18}
{Thompson}, S.~E., {Coughlin}, J.~L., {Hoffman}, K., {et~al.} 2018, \apjs, 235,
  38

\bibitem[{{Van Eylen} {et~al.}(2019){Van Eylen}, {Albrecht}, {Huang},
  {MacDonald}, {Dawson}, {Cai}, {Foreman-Mackey}, {Lundkvist}, {Silva Aguirre},
  {Snellen}, \& {Winn}}]{vaneylen2019}
{Van Eylen}, V., {Albrecht}, S., {Huang}, X., {et~al.} 2019, \aj, 157, 61

\bibitem[{{Volk} \& {Gladman}(2015)}]{Volk2015}
{Volk}, K., \& {Gladman}, B. 2015, \apjl, 806, L26

\bibitem[{{Weiss} \& {Marcy}(2014)}]{Weiss2014}
{Weiss}, L.~M., \& {Marcy}, G.~W. 2014, \apjl, 783, L6

\bibitem[{{Weiss} \& {Petigura}(2019)}]{weis19}
{Weiss}, L.~M., \& {Petigura}, E.~A. 2019, arXiv e-prints, arXiv:1908.05833

\bibitem[{{Weiss} {et~al.}(2018){Weiss}, {Marcy}, {Petigura}, {Fulton},
  {Howard}, {Winn}, {Isaacson}, {Morton}, {Hirsch}, {Sinukoff}, {Cumming},
  {Hebb}, \& {Cargile}}]{weis18}
{Weiss}, L.~M., {Marcy}, G.~W., {Petigura}, E.~A., {et~al.} 2018, \aj, 155, 48

\bibitem[{{Welsh} {et~al.}(2015){Welsh}, {Orosz}, {Short}, {Cochran}, {Endl},
  {Brugamyer}, {Haghighipour}, {Buchhave}, {Doyle}, {Fabrycky}, {Hinse},
  {Kane}, {Kostov}, {Mazeh}, {Mills}, {M{\"u}ller}, {Quarles}, {Quinn},
  {Ragozzine}, {Shporer}, {Steffen}, {Tal-Or}, {Torres}, {Windmiller}, \&
  {Borucki}}]{Welsh2015}
{Welsh}, W.~F., {Orosz}, J.~A., {Short}, D.~R., {et~al.} 2015, \apj, 809, 26

\bibitem[{{Wolff} {et~al.}(2012){Wolff}, {Dawson}, \&
  {Murray-Clay}}]{Wolff2012}
{Wolff}, S., {Dawson}, R.~I., \& {Murray-Clay}, R.~A. 2012, \apj, 746, 171

\bibitem[{{Wolfgang} {et~al.}(2016){Wolfgang}, {Rogers}, \&
  {Ford}}]{Wolfgang2016}
{Wolfgang}, A., {Rogers}, L.~A., \& {Ford}, E.~B. 2016, \apj, 825, 19

\bibitem[{{Xie} {et~al.}(2016){Xie}, {Dong}, {Zhu}, {Huber}, {Zheng}, {De Cat},
  {Fu}, {Liu}, {Luo}, {Wu}, {Zhang}, {Zhang}, {Zhou}, {Cao}, {Hou}, {Wang}, \&
  {Zhang}}]{xie16}
{Xie}, J.-W., {Dong}, S., {Zhu}, Z., {et~al.} 2016, Proceedings of the National
  Academy of Science, 113, 11431

\bibitem[{{Yoshinaga} {et~al.}(1999){Yoshinaga}, {Kokubo}, \&
  {Makino}}]{Yoshinaga1999}
{Yoshinaga}, K., {Kokubo}, E., \& {Makino}, J. 1999, \icarus, 139, 328

\bibitem[{{Zhou} {et~al.}(2007){Zhou}, {Lin}, \& {Sun}}]{Zhou2007}
{Zhou}, J.-L., {Lin}, D.~N.~C., \& {Sun}, Y.-S. 2007, \apj, 666, 423

\bibitem[{{Zhu}(2019)}]{zhu19}
{Zhu}, W. 2019, arXiv e-prints, arXiv:1907.02074

\bibitem[{{Zink} {et~al.}(2019){Zink}, {Christiansen}, \& {Hansen}}]{zink19}
{Zink}, J.~K., {Christiansen}, J.~L., \& {Hansen}, B. M.~S. 2019, \mnras, 483,
  4479

\end{thebibliography}

\end{document}